\documentclass[11pt,a4paper]{article}
\usepackage{jcappub}

\usepackage{enumitem}
\usepackage{accents}

\newcommand{\bdot}[1] {\overset{\,_{\mbox{\bf\large .}}}{#1}}
\newcommand{\bddot}[1] {\overset{\,_{\mbox{\bf\large ..}}}{#1}}
\newcommand\gortho{{\overset{\bullet}{g}}{}}
\newcommand\hortho{{\overset{\bullet}{h}}{}}

\newcommand\horthop{{\overset{p}{h}}{}}
\newcommand\eortho{{\overset{\bullet}{e}}{}}
\newcommand\Wortho{{\overset{\bullet}{\omega}}{}}

\newcommand\omegaLC{{\overset{\circ}{\omega}}{}}
\newcommand\DLC{{\overset{\ \circ}{D}}{}}

\newcommand\RLC{{\overset{\ \circ}{R}}{}}
\newcommand\GLC{{\overset{\ \circ}{G}}{}}

\title{Bianchi Type Cosmological Models in $f(T)$ Tele-parallel Gravity}

\author[a]{R. J.  {van den Hoogen}}
\emailAdd{rvandenh@stfx.ca}
\affiliation[a]{Department of Mathematics, Statistics, and Computer Science, St. Francis Xavier University, Antigonish, Nova Scotia, Canada, B2G 2W5}

\author[b]{A. A.  {Coley}}
\emailAdd{aac@mathstat.dal.ca}
\affiliation[b]{Department of Mathematics and Statistics, Dalhousie University, Halifax, Nova Scotia, Canada, B3H 3J5}

\author[c]{D. D.  {McNutt}}
\emailAdd{mcnuttdd@gmail.com}
\affiliation[c]{ Department of Mathematics and Statistics, UiT: The Arctic University of Norway, Tromsø, Norway, 9019}

\date{\today}

\abstract
{Symmetry assumptions on the geometrical framework have provided successful mechanisms to develop physically meaningful solutions to many problems.  In tele-parallel gravity, invariance of the frame and spin-connection under a group of motions defines an affine symmetry group.  Here, we assume there exists a three-dimensional group of affine symmetries acting simply transitively on a spatial hypersurface and that this group of symmetry actions defines our affine frame symmetry group.  We determine the general form of the co-frame and spin connection for each spatially homogeneous Bianchi type.  We then construct the corresponding field equations for $f(T)$ tele-parallel gravity. We show that if the symmetry group is of Bianchi type A ($I$, $II$, $VI_0$, $VII_0$, $VIII$ or $IX$) then there exists a co-frame/spin connection pair that is consistent with the antisymmetric part of the field equations of $f(T)$ tele-parallel gravity.   For those geometries having a Bianchi type B symmetry group ($IV$, $V$, $VI_h$, $VII_h$), we find that in general these geometries are inconsistent with the antisymmetric part of the $f(T)$ tele-parallel gravity field equations unless the theory reduces to an analog of General Relativity with a cosmological constant.  
}

\begin{document}
\maketitle


\section{Introduction}

\subsection{Introductory Remarks}

It is well known that the theory of General Relativity (GR) has been extremely successful in explaining a variety of gravitational phenomena.  Indeed, GR is in agreement with almost all experimental and observational tests to date \cite{Will:2014kxa}.  It is only on cosmological scales in which there appears to be any divergence from the predictions of GR; for example, the currently observed accelerated expansion of our universe.  Even then, with some modifications to the matter sector of the theory, the addition of cold dark matter and a cosmological constant, one arrives at a cosmological model that agrees extremely well with current observations \cite{WMAP:2012nax,Planck:2018vyg}.  While the $\Lambda CDM$ models work very well, the curious nature of both dark energy and dark matter makes us ask the question ``Can we explain these same effects geometrically using an alternative theory to GR?".

Investigating alternatives to GR is not a recent phenomena; even Einstein himself spent time on this venture \cite{Nojiri:2010wj,Nojiri:2005jg,Nojiri_Odintsov2006,Capozziello_DeLaurentis_2011,Clifton:2011jh,CANTATA:2021ktz}.  One class of alternative theories of gravity is tele-parallel Gravity \cite{Aldrovandi_Pereira2013,Krssak:2018ywd,Krssak_Saridakis2015,Obukhov_Pereira2003,Maluf2013,MuellerHoissen_Nitsch1983}.  Interestingly, there is even a tele-parallel theory of gravity that is dynamically equivalent to GR and therefore indistinguishable from GR via classical tests of gravity \cite{Aldrovandi_Pereira2013}.  The tele-parallel equivalent to General Relativity (TEGR), while dynamically equivalent, approaches gravity via a very different perspective than GR.  Further, TEGR can be framed as a gauge theory for the translation group, something that has  not yet been possible in GR.  The Lagrangian for TEGR is constructed from a scalar $T$ which is constructed from scalar invariants of the torsion.  This torsion scalar is constructed in a particular way so that there exists a relationship between it and the Ricci scalar of the Levi-Civita connection plus a divergence term. The TEGR has been well developed and clearly explained in a number of venues (see in particular \cite{Aldrovandi_Pereira2013}).

In TEGR, the torsion scalar $T$ is a particular linear combination of scalar invariants constructed from the torsion.  By considering arbitrary linear combinations of these same scalar invariants of the torsion, Hayashi and Shirafuji \cite{Hayashi:1979qx} developed an alternative tele-parallel theory of gravity which they called New General Relativity (NGR). NGR, an interesting theory of gravity in its own right, is not considered in this paper.   Another modification to the standard TEGR is a theory in which one considers functions of this same torsion scalar, $T$, in the Lagrangian \cite{Krssak:2018ywd,Krssak_Saridakis2015,Ferraro:2006jd,Ferraro:2008ey,Linder:2010py}.  This class of tele-parallel theories of gravity is colloquially known as $f(T)$ tele-parallel gravity and reduces to TEGR when $f(T)=T$.  This generalization of TEGR has been a popular alternative theory in recent decades.  A comprehensive review of large classes of tele-parallel gravity theories can be found in \cite{Cai_2015,Bahamonde:2021gfp}.  We shall consider $f(T)$ tele-parallel theories of gravity.

In $f(T)$ tele-parallel gravity there has been a challenge in determining the appropriate ansatz in which to begin any investigation into finding physically meaningful solutions.  Because we have both a spin connection $\omega^a_{~b}=\omega^a_{~b\mu}dx^\mu$ and a co-frame, $h^a=h^a_{~\mu}dx^\mu$ , the two inputs must be in some sense compatible with one another.  The typical example of this occurs when one assumes a diagonal co-frame in spherically symmetric spacetimes and a trivial spin-connection. A diagonal co-frame is defined so that the matrix $h^a_{~\mu}$ defining the co-frame  is diagonal. While this diagonal co-frame and spin connection pair is a simple initial ansatz, they are inconsistent in the sense that the antisymmetric part of the field equations immediately impose $f(T)=T$.  One should start the analysis with a co-frame/spin connection pair that are consistent; that is, the field equations, in particular the antisymmetric part of the field equations, impose no or minimal constraints on the function $f(T)$ which defines the theory. Even though we always have the gauge freedom to choose a proper co-frame in which $\omega^a_{~b} = 0$, we must carefully choose the corresponding proper co-frame that leads to a consistent system.  The gauge choice is on the pair of inputs, not on just the co-frame or the spin connection.

For cosmological purposes, in most of the literature to date, the geometrical framework (frame or co-frame plus spin connection) are fixed so that the resulting geometry is characterized by a $k=0$ spatially homogeneous and isotropic metric. Recently, employing the ideas of symmetry invariance of the geometrical framework through a $G_6$ group of symmetry motions,  the appropriate frame and spin connection ansatz which result in a curved $k=\pm 1$ spatially homogeneous and isotropic metric has been developed \cite{Hohmann:2019nat,Hohmann:2021ast,DAmbrosio:2021zpm,DAmbrosio:2021pnd,Coley:2022qug,Bahamonde:2021gfp}, thereby yielding the tele-parallel Robertson Walker (TRW) geometries.  Prior to these developments, one was required to employ complex valued frame ansatzes when $k=-1$ (for example), thereby limiting their general applicability. It is now of interest to expand this idea of symmetry invariance on the geometrical framework by investigating what happens when we drop the isotropy requirement and look at simply transitive $G_3$ groups of motions acting on three dimensional spatial hypersurfaces; i.e, the tele-parallel Bianchi affine geometries.


We begin this exposition with a perhaps slightly unconventional approach to developing the field equations of $f(T)$ tele-parallel gravity.  We use the Metric Affine Gauge (MAG) framework of Hehl et al. \cite{Hehl_McCrea_Mielke_Neeman1995} to determine the field equations.  We employ Lagrange multipliers to ensure the zero curvature and zero non-metricity constraints.  Thereby, ensuring a fully covariant development of the field equations.  We assume that the geometrical variables (i.e., the co-frame and spin connection) are invariant under a simply transitive $G_3$ group of motions and consequently we solve the antisymmetric part of the field equations.  Solutions of the antisymmetric part of the field equations when they exist, provide a consistent ansatz for the spin-connection that can be used in the symmetric part of the field equations.  Given the co-frame and the consistent spin connection, we write the symmetric part of the $f(T)$ tele-parallel gravity field equations for Bianchi type A geometries with a comoving perfect fluid. While Bianchi type A affine geometries are consistent with the $f(T)$ tele-parallel gravity field equations, Bianchi type B affine geometries are only consistent in special cases.

\subsection{Review of other work}

Spatially homogeneous but anisotropic cosmological models in $f(T)$ tele-parallel gravity have been investigated in the literature with limited success and generality. Bianchi type I cosmological models in $f(T)$ tele-parallel gravity are the most common spatially homogeneous and isotropic geometry analyzed in the literature
\cite{Rodrigues:2012qua,Rodrigues:2013iua,Rodrigues:2014xam,Sharif:2011bi,
Amir:2015wja,Fayaz:2014swa,Cai_2015,
Fayaz:2015yka,Skugoreva:2017vde,Skugoreva:2019bwt,
Tretyakov:2021cgb,Aslam:2013coa,Paliathanasis:2016vsw,Coley:2023ibm}. In \cite{Rodrigues:2012qua,Rodrigues:2014xam,Sharif:2011bi} the authors construct the function $f(T)$ based on some external assumptions on the matter source; for example, dust or radiation.  Alternatively, these same authors determined the effective equation of state for a given $f(T)$. Others \cite{Fayaz:2014swa,Fayaz:2015yka} investigated the possibility of constructing $f(T)$ from both holographic dark energy (HDE) and Ricci dark energy (RDE) arguments. A comprehensive review and exact solutions can be found in \cite{Cai_2015}.

For Bianchi type I geometries, the simple ansatz of a diagonal co-frame with a trivial spin connection is always consistent with the $f(T)$ tele-parallel gravity field equations. This means that a proper diagonal Bianchi type I co-frame is consistent with the field equations; this is sometimes referred to as a ``good'' co-frame \cite{Tamanini_Boehmer2012}. However, it must be noted, that the assumption of a trivial spin connection for Bianchi I spatially homogeneous affine geometries is not the only ansatz that is consistent with the field equations \cite{Coley:2023ibm}.  It is possible to construct special classes of Bianchi type I affine geometries in which the co-frame is still diagonal but the spin-connection is non-trivial.

Bianchi type III anisotropic affine geometries in $f(T)$ tele-parallel gravity have been analyzed in \cite{Rodrigues:2012qua,Rodrigues:2014xam,Cai_2015}.  With the assumptions of a diagonal co-frame and trivial spin connection, one of the field equations immediately reduces to the equivalent of $\partial_t(F'(T))=0$ \cite{Rodrigues:2012qua,Rodrigues:2014xam,Cai_2015}.  This implies that the $f(T)$ tele-parallel gravity theory becomes equivalent to TEGR (or GR) with a cosmological constant and a re-scaled matter coupling constant.  One can interpret this in the following way: the assumption of a diagonal co-frame with a trivial spin connection is inconsistent with the $f(T)$ tele-parallel gravity field equations when the affine geometry is Bianchi type III.  Either the co-frame should be non-diagonal or the spin connection should be non-trivial.  Colloquially, this means that a proper diagonal Bianchi type III co-frame is not a ``good'' co-frame. Is it even possible to find a co-frame/spin-connection pair for Bianchi type III geometries that does not lead immediately to TEGR (or GR) with a cosmological constant?

The most general approaches to the development of spatially homogeneous and anisotropic $f(T)$ tele-parallel geometries are found in \cite{Bahamonde:2021gfp,Hohmann:2023sto}.  Proper co-frame ansatzes in which the spin connection is trivial for a number of Bianchi types have been proposed in \cite{Bahamonde:2021gfp}. However, it is not clear whether these proper co-frame ansatzes can be utilized in general.  Indeed, even in the simplest case of a Bianchi type I in \cite{Bahamonde:2021gfp}, the proposed proper co-frame ansatz leads to field equations in $f(T)$ tele-parallel gravity that depend upon the spatial coordinates, a somewhat odd observation given the spatial homogeneous nature of the geometry.  This observed inconsistency is due to the fact that the authors of \cite{Bahamonde:2021gfp} use the symmetries of the metric (i.e., Killing symmetries) to define/restrict the geometry, but the field equations for $f(T)$ tele-parallel gravity are not invariant under these Killing symmetries. Therefore the resulting field equations may lead to some restrictions on the theory. Here, in this paper we use a different notion of symmetry.  Affine frame symmetries \cite{Coley:2019zld,Coley:2022qug,McNutt:2023nxm} are employed to define/restrict the geometry in which case the geometrical side of the $f(T)$ tele-parallel gravity field equations are also invariant under these same affine frame symmetries.

The recent paper \cite{Hohmann:2023sto} can be considered to be one of the first in which a much more general mathematical approach to constructing co-frame/spin connection ansatzes for tele-parallel geometries is developed. The technique employed in \cite{Hohmann:2023sto} is applied to three special classes of spatially homogeneous geometries, Bianchi types $II$, $III$, and $IX$, that have an additional fourth symmetry.  Hohmann \cite{Hohmann:2023sto} works in a proper frame, and determines the corresponding co-frame that is respectful of the four assumed symmetry assumptions.  The subsequent results contain six arbitrary functions of time in the solution to the co-frame, at least one of which can always be eliminated through a coordinate transformation. It is not yet clear how the definitions and approach employed in \cite{Hohmann:2023sto,Hohmann:2019nat} for example, relate to the ideas of affine frame symmetries proposed in \cite{Coley:2019zld,Coley:2022qug,McNutt:2023nxm}.  While the results in \cite{Hohmann:2023sto} are general and can be employed in any tele-parallel theory of gravity, the proper co-frames that are developed have not been tested for their consistency in the $f(T)$ tele-parallel gravity field equations, that is, it is not known whether the co-frames determined in \cite{Hohmann:2023sto} lead to a ``good'' proper co-frame or not.

To really test the cosmological fitness of $f(T)$ tele-parallel gravity theories requires that we are able to test cosmological observations against not just isotropic models, but also anisotropic models.  With this goal in mind, it is imperative to develop the framework to properly construct anisotropic spatially homogenous models in $f(T)$ tele-parallel gravity.  Without this solid foundation, it becomes very difficult to truly ascertain the feasibility of $f(T)$ tele-parallel gravity in cosmological settings.

\subsection{Notation}

We restrict the analysis to four dimensional differentiable manifolds with coordinates $x^\mu$.  The notation employed uses Greek indices $\{\mu,\nu,\dots\}$ to represent space-time coordinate indices and Latin indices $\{a,b,\dots\}$, represent frame or tangent-space indices. Upper case Latin indices $\{I,J,\dots\}$ represent the restricted set of tangent space indices $\{1,2,3\}$.  Particular values of tangent or cotangent space indices are indicated with a hat.  Lower case Latin indices $\{i,j,\dots\}$ represent the restricted set of coordinate indices $\{1,2,3\}$.  Round brackets surrounding indices represents symmetrization, while square brackets represents anti-symmetrization. Any underlined index is not included in the symmetrization. Quantities with a solid bullet indicate that a \emph{Orthonormal Gauge} choice has been made.  Quantities with an over-circle are calculated using the Levi-Civita connection, $\omegaLC^a_{\phantom{a}b}$. Derived geometric quantities such as curvature or torsion depend on the co-frame and spin-connection or both.  This dependence is indicated where necessary via square brackets, for example, the curvature two-form of the Levi-Civita connection is denoted as $R^a_{\phantom{a}b}[\omegaLC^c_{\phantom{c}d}]=\RLC^a_{\phantom{a}b}$.  The orthonormal metric has the form $\gortho_{ab}=\mbox{Diag}[-1,1,1,1]$.
The frame basis is denoted as $e_a$ with the corresponding co-frame basis $h^a$ where $e_a\rfloor h^b = \delta^a_b$.  We define the volume four form $\eta={}^*(1)$ as the dual to the unit zero form and through continued inner products with frame $e_a$ we are able to define $\eta_a=e_a\rfloor\eta = {}^*h_a$, $\eta_{ab}=e_a\rfloor e_b\rfloor \eta = {}^*(h_a\wedge h_b)$ and $\eta_{abc}=e_a\rfloor e_b\rfloor e_c\rfloor \eta = {}^*(h_a\wedge h_b \wedge h_c)$.  The canonical frame basis for each Bianchi type will be denoted as $\mathcal{E}_I$ with corresponding co-frame basis $\mathcal{H}^I$.


\section{Basics of Tele-parallel Gravity}\label{BasicTEGR}

\subsection{Geometry and Gauge Choices}

The Metric Affine Gauge theory (see \cite{Hehl_McCrea_Mielke_Neeman1995}) provides a suitable framework to determine the field equations describing large classes of tele-parallel (and other) theories of gravity.  One begins with a four dimensional differentiable manifold $M$ with symmetric metric $g_{ab}$, co-frame one-form basis $h^a$ and spin connection one form $\omega^a_{~b}$. In a general, the set of geometrical quantities
\begin{equation}
\left\{ g_{ab},h^a,\omega^{a}_{\phantom{a}b} \right\} \label{original_geometry}
\end{equation}
defines the geometry. The three geometrical quantities are independent and only become constrained as we entertain different geometrical or physical assumptions.

The corresponding field strengths associated with the metric, co-frame and spin connection are; the symmetric non-metricity one-form ($Q_{ab}$), the torsion two form ($T^a$), and the curvature two-form ($R^a_{\phantom{a}b}$) defined as
\begin{eqnarray}
Q_{ab}&=&-Dg_{ab},\quad T^a=Dh^a, \quad R^a_{\phantom{a}b}=d\omega^a_{\phantom{a}b}+\omega^a_{\phantom{a}c}\wedge\omega^c_{\phantom{a}b}.
\end{eqnarray}
The associated Bianchi identities are
\begin{eqnarray}
DQ_{ab}             &=& 2R_{(ab)},\label{Bianchi1}\\
DT^a                &=& R^{a}_{\phantom{a}c} \wedge h^c, \label{Bianchi2}\\
DR^a_{\phantom{a}b} &=& 0\label{Bianchi3},
\end{eqnarray}
which yield additional differential constraints between the field strengths.

It is widely accepted that the laws of gravitational physics be invariant under any change of coordinates ({\em General Covariance}) and be invariant under arbitrary changes in the frame of reference ({\em Principle of Relativity}) \cite{Ortin:2004ms,Hehl_McCrea_Mielke_Neeman1995}.  Consequently with these assumptions, the resulting field equations derived from any given Lagrangian will have to transform covariantly under both local $GL(4)$ gauge transformations of the co-frame, metric, and affine connection fields, and transform covariantly under general coordinate transformations.
With the \emph{Principle of Relativity} and the consequent covariance of the field equations under local $GL(4)$ gauge transformations, one is permitted to judiciously simplify aspects of the calculations via a well-chosen gauge choice. For example, a convenient choice is to employ the \emph{Orthonormal Gauge} in which the metric becomes $g_{ab}=\gortho_{ab}=\mbox{Diag}[-1,1,1,1]$.   We note that each of these invariance assumptions also imply the existence of an associated Noether identity.

It is worth mentioning at this point that the Bianchi identities, equations \eqref{Bianchi1}-\eqref{Bianchi3}, for tele-parallel gravity in which $R^a_{\phantom{a}b}=0$ and $Q_{ab}=0$, are just that, identities.  There is no new information or constraints obtained from the Bianchi identities.  This is, in direct contrast to GR, where the torsion, $T^a$ and non-metricity, $Q_{ab}$ are trivial, in which case the Bianchi identities \eqref{Bianchi1}-\eqref{Bianchi3}, yield the non-trivial algebraic constraint on the curvature, $R_{(ab)}$ and $R^a_{\phantom{a}b}\wedge h^b=0$.  So while torsion is zero in GR, the absence of torsion does impart indirectly a nontrivial effect.

\subsection{The Matter Field Lagrangian}

In general the matter fields may be described using scalar, tensor or spinor-valued forms of any rank. For illustrative purposes, it is sufficient to consider a scalar field $\Phi$ and a vector valued $1$-form field $\Psi^a$ as the source fields for the matter.  The matter field Lagrangian $4$-form, $L_{Matt}$, is in general a function of the field variables $g_{ab}$, $h^a$, $\omega^a_{\phantom{a}b}$, $\Phi$, $\Psi^a$ and their corresponding exterior derivatives.
However, assuming the {\it Principle of Relativity}, the spin connection, $\omega^a_{\phantom{a}b}$ can only enter the matter field Lagrangian indirectly through covariant derivatives of the various field variables, $Q_{ab},T^a,R^a_{\phantom{a}b}$, $D\Phi$ and $D\Psi^a$.
With the additional assumption of a {\em Minimal Coupling} between the matter fields and the gauge fields \cite{Ciufolini_Wheeler1995,vonderHeyde1975,Hehl_McCrea_Mielke_Neeman1995,Hehl_vonderHeyde_Kerlick_Nester1976}, the matter field Lagrangian is independent of the derivatives $dg_{ab},dh^a,d\omega^a_{\phantom{a}b}$. Therefore, assuming the {\em Principle of Relativity} together with {\em Minimal Coupling} the matter field Lagrangian has the form
\begin{equation}
L_{Matt}=L_{Matt}(g_{ab},h^a,\Phi,D\Phi,\Psi^a,D\Psi^a).\label{Matter_Lagrangian}
\end{equation}
The corresponding metrical, canonical, and the hyper-momentum matter currents  \cite{Hehl_McCrea_Mielke_Neeman1995} are then
\begin{eqnarray}
\sigma^{ab} &\equiv& \frac{\delta L_{Matt}}{\delta g_{ab}} = \frac{\partial L_{Matt}}{\partial g_{ab}},\\
\Sigma_a    &\equiv& \frac{\delta L_{Matt}}{\delta h^a} = \frac{\partial L_{Matt}}{\partial h^a},\\
\Delta^a_{\phantom{a}b}&\equiv&\Psi^a \wedge \left(\frac{\partial L_{Matt}}{\partial(D \Psi^b)}\right),
\end{eqnarray}
where the spin matter current $\tau_{ab}$ is defined to be the antisymmetric part of the hyper-momentum current $\tau_{ab}=\Delta_{[ab]}$. The invariance of the matter field Lagrangian under general linear transformations of the co-frame yields the Noether conservation identity for the matter fields
\begin{equation}
D\Delta^a_{~b}+h^a\wedge \Sigma_b - g_{bc}\sigma^{ac} + \Psi^a\wedge \frac{\delta L_{Matt}}{\delta\Psi^b} =0. \qquad \mbox{[NOETHER M]}\label{NOETHER-M}
\end{equation}
Assuming that the field equation for $\Psi^a$ is satisfied (i.e., on shell), then this conservation identity yields a relationship between the three various matter currents.  Indeed, if the matter Lagrangian only contains a scalar field $\Phi$, then $\Delta^a_{~b}=0$ and
\begin{equation}
h^{(a}\wedge \Sigma^{b)} = \sigma^{ab} \qquad \mbox{and} \qquad h^{[a}\wedge \Sigma^{b]} = 0 \label{NOETHER-M2}
\end{equation}
which shows the equivalence between the metrical energy momentum current $\sigma^{ab}$ and the canonical energy momentum current $\Sigma^a$ when there is only scalar matter.  Alternatively, one clearly sees how the assumption of the {\it Principle of Relativity} leads to a symmetric canonical energy momentum current via the corresponding Noether identity.

\subsection{The Gauge Field Lagrangian}

The Lagrangian $4$-form for the gauge fields, $V_{gauge}$, can be a function of the field variables $g_{ab},h^a,\omega^a_{\phantom{a}b}$ and their corresponding exterior derivatives. Assuming the \emph{Principle of Relativity}, implies that the spin connection, $\omega^a_{\phantom{a}b}$, enters the gauge field Lagrangian only through $Q_{ab},T^a,R^a_{\phantom{a}b}$,
\begin{equation}
V_{gauge}=V_{gauge}(g_{ab},h^a,Q_{ab},T^a,R^{a}_{~b}).\label{Gauge_Lagrangian}
\end{equation}
The corresponding gauge field momenta \cite{Hehl_McCrea_Mielke_Neeman1995} are defined
\begin{eqnarray}
M^{ab} &\equiv&2\frac{\partial V_{gauge}}{\partial dg_{ab}}=-2\frac{\partial V_{gauge}}{\partial Q_{ab} },\\
H_a &\equiv&-\frac{\partial V_{gauge}}{\partial dh^a}=-\frac{\partial V_{gauge}}{\partial T^a},\label{gauge_FM2}\\
H^{a}_{\phantom{a}b}&\equiv&-\frac{\partial V_{gauge}}{\partial (d\omega^b_{\phantom{a}a})}=-\frac{\partial V_{gauge}}{\partial R^b_{\phantom{a}a}},\label{gauge_FM3}
\end{eqnarray}
and the metrical energy momentum, canonical energy momentum and the hyper-momentum associated with the gauge fields \cite{Hehl_McCrea_Mielke_Neeman1995} are
\begin{eqnarray}
m^{ab} &\equiv&2\frac{\partial V_{gauge}}{\partial g_{ab}},\\
E_a&\equiv&\frac{\partial V_{gauge}}{\partial h^a},\label{gauge_EM2}\\
E^a_{\phantom{a}b}&\equiv&\frac{\partial V_{gauge}}{\partial \omega^b_{\phantom{a}a}} =-h^a \wedge H_{b}-g_{bc}M^{ac}.\label{gauge_EM3}
\end{eqnarray}

Due to the assumption of the {\em Principle of Relativity}, the corresponding Noether conservation identity for the gauge fields is
\begin{equation}
-R^a_{~c}\wedge H^c_{~b}+R^c_{~b}\wedge H^a_{~c}-T^a\wedge H_b+h^a\wedge E_b+Q_{bc}M^{ac}-g_{bc}m^{ac} =0. \mbox{[NOETHER G]}\label{NOETHER-G}
\end{equation}

\subsection{The Action and Field Equations for Tele-parallel Gravity}

Tele-parallel Gravity in which the nonmetricity and curvature are assumed to be zero can be achieved within the Metric Affine Gauge theory of gravity via Lagrange multipliers.  We introduce the Lagrange multiplier $\mu^{ab}$ as a symmetric $3$ form and the Lagrange multiplier $\nu_a^{\phantom{a}b}$ as a $2$ form. Therefore, the action for tele-parallel gravity in which the matter is minimally coupled to the gauge fields is
\begin{equation}
 S=\int \left[V_{gauge}(g_{ab},h^a,Q_{ab},T^a,R^{a}_{\phantom{a}b}) +\frac{1}{2}Q_{ab}\wedge\mu^{ab} +R^{a}_{\phantom{a}{b}}\wedge \nu_a^{\phantom{a}{b}} +L_{Matt}(g_{ab} h^a,\Phi,D\Phi,\Psi^a,D\Psi^a)\right] . \label{action}
\end{equation}
Using the Bianchi identities (\ref{Bianchi1})-(\ref{Bianchi3}), it can be shown that above action is invariant under the redefinitions of the Lagrange multipliers,
\begin{eqnarray}
\mu^{ab}\to\mu^{ab}+D\xi^{ab},\\
\nu_a^{\phantom{a}b} \to \nu_a^{\phantom{a}b} +D\chi^b_{\phantom{a}a}-\xi^{b}_{\phantom{a}a},
\end{eqnarray}
where $\xi^{ab}=\xi^{ba}$ is an arbitrary symmetric $2$ form, and $\chi^a_{\phantom{a}b}$ is an arbitrary $1$-form \cite{Obukhov_Pereira2003}.  Therefore it is impossible to determine the Lagrange multipliers uniquely.

Variation of this action with respect to the Lagrange multipliers $\mu^{ab}$ and  $\nu_a^{\phantom{a}b}$ and the field variables $g_{ab},h^a,\omega^a_{\phantom{a}b},\Phi,\Psi^a$, \cite{Hehl_McCrea_Mielke_Neeman1995,Obukhov_Pereira2003} yields the following set of field equations
\begin{eqnarray}
Q_{ab}=0, &&\qquad\mbox{[NONMETRICITY]}\label{NONMETRICITY}\\
R^a_{\phantom{a}b}=0, &&\qquad\mbox{[CURVATURE]}\label{CURVATURE}\\
DM^{ab}-m^{ab}-D\mu^{ab}=\sigma^{ab}, &&\qquad\mbox{[ZEROTH]}\label{ZEROTH}\\
DH_a-E_a=\Sigma_a, &&\qquad\mbox{[FIRST]}\label{FIRST}\\
DH^a_{\phantom{a}b}-E^a_{\phantom{a}b}-g_{bc}\mu^{ac}-D\nu_b^{\phantom{a}a}=\Delta^a_{\phantom{a}b}, &&\qquad\mbox{[SECOND]}\label{SECOND}\\
\frac{\delta L_{Matt}}{\delta\Phi}=0, &&\qquad\mbox{[MATTER\ 1]} \label{MATTER1}\\
\frac{\delta L_{Matt}}{\delta\Psi^a}=0. &&\qquad\mbox{[MATTER\ 2]} \label{MATTER2}
\end{eqnarray}
We follow the same labelling as was used in \cite{Hehl_McCrea_Mielke_Neeman1995}.

Given that the nonmetricity is zero, we can raise and lower indices with ease.  The symmetric part of [SECOND] will give us an expression for the Lagrange multiplier $\mu^{ab}$. Using this expression for $\mu^{ab}$, and using the expression [NOETHER-G] for $m^{ab}$, as well as the fact that the curvature and non-metricity are zero,  and substituting results into equation [ZEROTH], one obtains the Noether identity [NOETHER-M]. Therefore, the [ZEROTH] equation can be eliminated without loss of generality.  The field equations that encode the dynamics are the [FIRST] equation, the antisymmetric part of [SECOND] and the matter field equations [MATTER\ 1] and [MATTER\ 2].

Recall [ZEROTH] is redundant and [SYM-SECOND] gives an expression for the Lagrange multiplier $\mu^{ab}$.  Indeed, [ASYM-SECOND] is not necessary to determine dynamics either, since it can be interpreted as determining part of the derivative of the Lagrange multiplier $\nu_{ab}$.

\section{$f(T)$ Tele-parallel Gravity}

The field equations (\ref{NONMETRICITY}--\ref{MATTER2}) together with the Noether identities in equations (\ref{NOETHER-M}) and (\ref{NOETHER-G}) are valid for broad classes of tele-parallel theories of gravity, including TEGR \cite{Aldrovandi_Pereira2013,Obukhov_Pereira2003}, $f(T)$ \cite{Krssak_Saridakis2015,Krssak:2018ywd}, New General Relativity \cite{Hayashi_Shirafuji1979,Hayashi_Shirafuji1982} and various generalizations \cite{Bahamonde:2016kba}.  We are interested in the class of $f(T)$ tele-parallel theories of gravity (which include TEGR as a special subcase) where $T$ is a scalar constructed from the torsion.

\subsection{The Torsion Scalar}

The torsion scalar $T$ is constructed from a linear combination of the following quadratic scalars
\begin{eqnarray*}
T_{TEN}&=&{}^*\left({}^{(1)}T^a\wedge{}^*{}^{(1)}T_a\right),\\
T_{TRA}&=&{}^*\left({}^{(2)}T^a\wedge{}^*{}^{(2)}T_a\right),\\
T_{AXI}&=&{}^*\left({}^{(3)}T^a\wedge{}^*{}^{(3)}T_a\right),
\end{eqnarray*}
which are the irreducible parts of the torsion that are invariant under the general linear group \cite{Hehl_McCrea_Mielke_Neeman1995}. See Appendix \ref{Torsion-Appendix} for details on the decomposition.
The torsion scalar $T$ is
\begin{eqnarray}
T &=& -T_{TEN}+2T_{TRA}+\frac{1}{2} T_{AXI},\label{torsion_scalar1}\\
  &=&{}^*(T^a\wedge {}^*S_a),
\end{eqnarray}
where we define
\begin{equation}
S_a= -{}^{(1)}T_a +2{}^{(2)}T_a+\frac{1}{2}{}^{(3)}T_a.
\end{equation}
Interestingly, both the torsion $T^a$ and the Hodge dual of $S_a$ can be expressed elegantly in terms of the contorsion one-form $K^a_{~b}$ as
\begin{equation}
T^a=K^a_{~b} \wedge h^b, \qquad\mbox{and}\qquad {}^*S_a=-\frac{1}{2}K^{cd}\wedge \eta_{acd}.
\end{equation}
which illustrates the naturalness of the coefficients $(-1,2,\frac{1}{2})$ of $T_{TEN}$, $T_{TRA}$ and $T_{AXI}$ in the definition of the torsion scalar in equation \eqref{torsion_scalar1} (see \cite{Obukhov:2006gea} and Appendix \ref{contorsion appendix} for additional details on the contorsion and the relationship between the torsion scalar and the Ricci scalar). The torsion scalar can also be represented purely in terms of the contorsion as
\begin{equation}
T={}^*\left(K^a_{\phantom{a}c}\wedge K^{bc}\wedge \eta_{ab}\right).
\end{equation}

\subsection{The Gauge Potential for $f(T)$ tele-parallel gravity}

The gauge potential 4-form, $V_{gauge}(g_{ab},h^a,T^a)$, for the class of $f(T)$ tele-parallel gravity theories is
\begin{equation}
V_{gauge}=-\frac{1}{2\kappa}f(T)\,\eta\nonumber
\end{equation}
where $\kappa=8\pi G$ and $\eta = {}^*1$ is the volume 4-form.   The gauge potential depends on an arbitrary twice differentiable function $f$ of the torsion scalar $T$. A negative is included so that when $f(T)=T$, the potential reduces to TEGR.

\subsection{The Field Equations}

Given our assumptions on the nature of $V_{gauge}(g_{ab},h^a,T^a)$ for tele-parallel gravity, we have
\begin{eqnarray}
H_a&=&-\frac{1}{\kappa}f'(T){}^*S_a,\\
E_a&=& -\frac{1}{2\kappa}f(T) e_a\rfloor\eta - \frac{1}{\kappa} f'(T) e_a\rfloor T^b \wedge {}^*S_b,\\
H^{a}_{\phantom{a}b}&=&0,\\
E^a_{\phantom{a}b}&=&-h^a\wedge H_b.
\end{eqnarray}

The [FIRST] field equation for $f(T)$ tele-parallel gravity can be expressed as
\begin{eqnarray}
-f''(T)dT\wedge {}^*S_a-f'(T)\Bigl(D\,{}^*S_a-e_a\rfloor T^b\wedge {}^*S_b\Bigr)+\frac{1}{2}f(T)\eta_a &=&\kappa \Sigma_a.
\end{eqnarray}
One can also express this [FIRST] equation in terms of the Riemman curvature 2-form for the Levi-Civita connection, $\RLC^{a}_{\phantom{a}b}$, as
\begin{equation}\label{first_2}
-f''(T)dT\wedge {}^*S_a - f'(T)\Bigl(\frac{1}{2}\RLC^{cd}\wedge \eta_{acd}+\frac{1}{2}T\eta_a\Bigr)+\frac{1}{2}f(T)\eta_a = \kappa \Sigma_a.
\end{equation}
Alternatively, by taking the wedge product of equation \eqref{first_2} with the one-form $h_b=g_{bc}h^c$ and then symmetrizing and anti-symmetrizing over the tangent space indices $a$ and $b$ we obtain respectively the symmetric and anti-symmetric parts of equation \eqref{first_2}
\begin{eqnarray}
-f''(T)\,dT\wedge {}^*S_{(a}\wedge h_{b)} +f'(T)\GLC_{ab}\eta-\frac{1}{2}f'(T)Tg_{ab}\eta+\frac{1}{2}f(T)g_{ab}\eta &=&\kappa \Sigma_{(a} \wedge h_{b)}, \label{sym_first1}\\
-f''(T)\,dT\wedge {}^*S_{[a}\wedge h_{b]} &=& \kappa \Sigma_{[a} \wedge h_{b]},
\end{eqnarray}
where $\GLC_{ab}$ is the Einstein tensor associated with the Levi-Civita connection.

The constraints $Q_{ab}=0$ and $R^a_{~b}=0$ together with the field equations, [SYM FIRST], [ASYM FIRST], [MATTER\ 1] and [MATTER\ 2] constitute the full set of field equations for $f(T)$ tele-parallel gravity
\begin{eqnarray*}
-f'(T)\Bigl(D{}^*S_{(a}-e_{(a}\rfloor T^c\wedge {}^*S_{\underline{c}}\Bigr)\wedge h_{b)} +\frac{1}{2}f(T)g_{ab}\eta& \phantom{=\kappa \Sigma_{(a} \wedge h_{b)}}  \nonumber\\
-f''(T)dT\wedge {}^*S_{(a}\wedge h_{b)} &=\kappa \Sigma_{(a} \wedge h_{b)} , &\qquad\mbox{[SYM-FIRST]}\\
-f''(T)\,dT\wedge {}^*S_{[a}\wedge h_{b]}&=\kappa \Sigma_{[a} \wedge h_{b]} , &\qquad\mbox{[ASYM-FIRST]}\\
 \frac{\delta L_{Matt}}{\delta\Phi}&=0, &\qquad\mbox{[MATTER\ 1]}\\
 \frac{\delta L_{Matt}}{\delta\Psi^a}&=0. &\qquad\mbox{[MATTER\ 2]}
\end{eqnarray*}


\section{Bianchi Cosmologies in $f(T)$ Tele-parallel Gravity}

\subsection{Affine Frame Symmetries}
An affine frame symmetry without isotropies is one in which there exists a pair $ (h^a, \omega^a_{~b})$ where both the co-frame and the spin-connection are invariant under the symmetry action, that is
\begin{equation}
\mathcal{L}_{\zeta}(h^a)=0, \qquad \qquad \mathcal{L}_{\zeta}(\omega^a_{\phantom{a}b})=0,\label{Affine_frame_symmetry1}
\end{equation}
where $\zeta$ is the vector generator(s) of the assumed symmetry \cite{Coley:2019zld,Fonseca-Neto:1992xln}.
We shall assume the existence of a $G_3$ group of symmetries acting simply transitively on three dimensional spatial hyper-surfaces and that this $G_3$ group defines the affine frame symmetries.  Let $\zeta_I$, $I\in \{1,2,3\}$, be a basis for the Lie algebra associated with the generators of this $G_3$ group of motions.

As we have complete freedom to choose a frame basis, we shall select a basis adapted to the assumed symmetries.  Let $n$ be a unit vector field normal to the orbits of the $G_3$ group of motions. Consequently $n$ is tangent to a geodesic congruence and we can choose the affine parameter along this geodesic as one of the coordinates, say $t$, and define $e_{\hat 0}=n=\partial_t$ or equivalently $h^{\hat 0}=dt$. An orthonormal spatial triad, $\eortho_I$, can be chosen through a time dependent linearly independent combination of the reciprocal group generators, $\mathcal{E}_J(x^i)$ in which case
\begin{equation}
\gortho_{ab}=Diag[-1,1,1,1] \quad \mbox{and} \quad \eortho_I=(M^{-1})^J_{~I}(t) \mathcal{E}_J(x^i)
\end{equation}
These $\mathcal{E}_J(x^i)$ are referred to as the {\it canonical} vector basis and expressions in local coordinates for each Bianchi type can be found in \cite{Stephani:2003tm}. Therefore, the symmetry adapted orthonormal co-frame $\hortho^a$, (determined via $ \eortho_a \rfloor \hortho^b = \delta^b_a$), can be chosen without loss of generality as
\begin{equation}
\hortho^a=[dt,M^I_{~J}(t)\mathcal{H}^J(x^i)]  \label{initial-frame}
\end{equation}
where $\mathcal{H}^J(x^i)$ are the corresponding {\it canonical} one-form basis \cite{Stephani:2003tm} and $M^I_{~J}(t)\in GL(3)$. These spatially homogeneous geometries are classified by the nature of the structure constants $\mathcal{C}^J=d\mathcal{H}^J$ into nine different Bianchi classes. (See Appendix \ref{canonical-coordinates} for explicit expressions for the canonical co-frame basis for each Bianchi type). The remaining spatially homogeneous geometry, not considered here, is when a $G_4$ group of symmetries acts multiply transitively on three dimensional spatial hyper-surfaces.  Consequently, these geometries have a one dimensional isotropy group, and require different techniques \cite{Coley:2022qug,McNutt:2023nxm}.

To simplify some aspects of computations later, we define $\mathcal{H}^{\hat{0}}=dt$, and define the matrix
\begin{equation}
M^a_{~c}(t) = \left[ \begin{array}{c|c}
   1 & 0 \\
   \hline
   0 & M^I_{~J}(t)
   \end{array}
   \right].
\end{equation}

With our chosen orthonormal basis, and given that $\mathcal{L}_{\zeta_I}(\omega^a_{\phantom{a}b})=0$ for $I\in\{1,2,3\}$, the components $\omega^a_{~bc}$ are functions of $t$ only; in which case
\begin{align}
\Wortho^a_{~b}&=\omega^a_{~b\hat{0}}(t)\,dt + \omega^a_{~bI}(t)\,\hortho^I(x^i) \label{ortho_omega} \\
              &=\omega^a_{~b\hat{0}}(t)\,\mathcal{H}^{\hat{0}} + \widetilde{\omega}^a_{~bI}(t)\,\mathcal{H}^I(x^i)\label{ortho_omega2}
\end{align}
in terms of the canonical co-frame basis, where we define $\widetilde{\omega}^a_{~bI}(t) = \omega^a_{~bJ}(t)M^J_{~I}(t)$.
Since we chose an orthonormal co-frame, there exists a matrix $\Lambda^a_{~b}(x^\mu) \in SO^+(1,3)$ such that the spin connection is
\begin{equation}
\Wortho^a_{\phantom{a}b}= (\Lambda^{-1})^{a}_{~c}d\Lambda^c_{~b}.   \label{omega_Lambda}
\end{equation}
In theory, we can always apply a Lorentz transformation to (\ref{initial-frame}) to bring us to a \emph {Proper Orthonormal} frame
\begin{equation}
\hortho^a\to \horthop^a=\Lambda^a_{~b}\hortho^b,\qquad \Wortho^{a}_{~b}\to 0^{a}_{~b},\qquad \gortho_{ab}\to \gortho_{ab},
\end{equation}
but we shall not impose this gauge restriction at this time. Alternatively, for each Bianchi type, we will solve for the explicit form of $\Wortho^a_{~b}$ from \eqref{omega_Lambda} in terms of the six Lorentz parameter functions that completely describe the Lorentz transformation.    Then, we will determine the conditions, if any, the $f(T)$ tele-parallel field equations place on the nature of $\Wortho^a_{~b}$.

\subsection{General Spin Connection for Bianchi Tele-parallel Geometries}\label{sec:spin}

What is the form of the general spin connection that can be employed in simply transitive spatially homogeneous tele-parallel geometries?  In \cite{Coley:2022aty} an algorithm is developed to study tele-parallel geometries with a single affine symmetry.  In a subsequent paper, the algorithm is generalized to Bianchi type I tele-parallel geometries \cite{Coley:2023ibm}.  In \cite{Coley:2023ibm}, the general form of the spin connection is computed in terms of three complex valued one-forms defined as
\begin{subequations}\label{DES_for_Lorentz}
\begin{align}
\Theta   &\equiv  (A^{-1}A_{,\mu} -  i \theta_{,\mu} - 2\bar{E}B_{,\mu})\,dx^\mu,\\
\Psi^{I}  &\equiv A^{-1}e^{i\theta}(\bar{E}_{,\mu}-\bar{E}B_{,\mu}\bar{E})\,dx^\mu,\\
\Psi^{II}  &\equiv A e^{-i\theta}B_{,\mu}\,dx^\mu,
\end{align}
\end{subequations}
where  $A(t,x^j)$ and $\theta(t,x^j)$ are real valued functions parameterizing the boost and spin of a general Lorentz transformation.  The complex valued functions $E(t,x^j)$ and $B(t,x^j)$ represent the two different possible null rotations.  See \cite{Coley:2022aty,Coley:2023ibm} for details.  While the computations in \cite{Coley:2022aty,Coley:2023ibm} were completed in a null frame, it is straightforward to transform to an orthonormal frame.  The general spin connection one-form in an orthonormal frame can be represented as
\begin{equation}
\Wortho^{a}_{~b}=\begin{bmatrix}
0                       & Re(\Theta)       & Re(\Psi^{I}+\Psi^{II})\quad   &  -Im(\Psi^{I}-\Psi^{II})  \\
Re(\Theta)              & 0                & Re(\Psi^{I}-\Psi^{II})   &  -Im(\Psi^{I}+\Psi^{II})  \\
Re(\Psi^{I}+\Psi^{II})  & -Re(\Psi^{I}-\Psi^{II}) \quad & 0 &Im(\Theta)  \\
-Im(\Psi^{I}-\Psi^{II})\quad & Im(\Psi^{I}+\Psi^{II})  & -Im(\Theta) &0
\end{bmatrix}
\label{gen_spin_ortho}
\end{equation}
where $Re(\ )$ and $Im(\ )$ indicate the real and imaginary parts of their respective arguments.

Now comparing equation \eqref{ortho_omega2} with equation \eqref{gen_spin_ortho} we obtain
\begin{subequations}\label{gg}
  \begin{align}
  \Theta &=  \Big(\widetilde{\omega}^{\hat{0}}_{~\hat{1}c}(t)+i\widetilde{\omega}^{\hat{2}}_{~\hat{3}c}(t)\Big)\mathcal{H}^c(x^j), \\
  \Psi^{I} &= \frac{1}{2}\Bigg( \left(\widetilde{\omega}^{\hat{0}}_{~\hat{2}c}(t)+\widetilde{\omega}^{\hat{1}}_{~\hat{2}c}(t)\right)
            -i\left(\widetilde{\omega}^{\hat{0}}_{~\hat{3}c}(t)+\widetilde{\omega}^{\hat{1}}_{~\hat{3}c}(t) \right)\Bigg)\mathcal{H}^c(x^j), \\
 \Psi^{II} &= \frac{1}{2}\Bigg(\left(\widetilde{\omega}^{\hat{0}}_{~\hat{2}c}(t)-\widetilde{\omega}^{\hat{1}}_{~\hat{2}c}(t)\right)
            +i\left(\widetilde{\omega}^{\hat{0}}_{~\hat{3}c}(t)-\widetilde{\omega}^{\hat{1}}_{~\hat{3}c}(t) \right)\Bigg)\mathcal{H}^c(x^j).
 \end{align}
 \end{subequations}
Expressing the components of the one-form $\Theta$ as $\Theta = \Theta_c(t) h^c(x^j)=\widetilde\Theta_c(t) \mathcal{H}^c(x^j)$ where $\widetilde{\Theta}_c(t)=\Theta_c(t) (M^{-1})^c_{~d}(t)$ and defining similarly $\widetilde{\Psi}^I(t)$ and $\widetilde{\Psi}^{II}(t)$ we obtain using equation \eqref{DES_for_Lorentz} the following system of differential equations
\begin{subequations}\label{DES_for_Lorentz2}
\begin{align}
\widetilde{\Theta}_c(t) \mathcal{H}^c_{~\mu}(x^j)  &=(A^{-1}A_{,\mu} -  i \theta_{,\mu} - 2\bar{E}B_{,\mu})   ,\\
\widetilde{\Psi}^{I}_c(t) \mathcal{H}^c_{~\mu}(x^j) &=A^{-1}e^{i\theta}(\bar{E}_{,\mu}-\bar{E}B_{,\mu}\bar{E})   ,\\
\widetilde{\Psi}^{II}_c(t) \mathcal{H}^c_{~\mu}(x^j)&=   A e^{-i\theta}B_{,\mu} .
\end{align}
\end{subequations}
We note that the system of differential equations is non-homogeneous in general -- that is there typically exist non-trivial source functions.  However, the source functions are separable functions of $t$ and $x^j$.  To find a solution, we propose a general separable solution of the form
\begin{subequations}\label{solutionasatz}
  \begin{align}
  A(t,x^j)e^{-\i\theta(t,x^j)} &= \alpha(t)e^{\lambda(x^j)},\\
  \bar{E}(t,x^j)  &= \bar\eta(t)e^{\lambda(x^j)},\\
  B(t,x^j)  &= \beta(t)e^{-\lambda(x^j)},
\end{align}
\end{subequations}
where $\alpha(t)$, $\eta(t)$ and $\beta(t)$ are complex valued functions of $t$, and $\lambda(x^j)$ is a complex valued function of the spatial coordinates $x^j$. This ansatz for the solution will yield a general solution provided that there are six real-valued arbitrary functions in the solution, else we find a special solution.  Substituting the general solution ansatz into equation \eqref{DES_for_Lorentz2} and expanding the one forms $\Theta$, $\Psi^I$ and $\Psi^{II}$ we find
\begin{subequations}
  \begin{align}
  \widetilde{\Theta}_0(t)\,dt +\widetilde{\Theta}_J(t)H^J_{~i}(x^j)\,dx^i &= \left(\frac{\dot{\alpha}}{\alpha}-2\bar\eta\dot\beta\right)\,dt
           +\left(1+2\bar\eta\beta\right)\frac{d\lambda}{dx^i}\,dx^i,\\
  \widetilde{\Psi}^I_0(t)\,dt +\widetilde{\Psi}^I_J(t)H^J_{~i}(x^j)\,dx^i &=
  \left( \dot{\bar\eta}-\bar\eta^2\dot\beta\right)\alpha^{-1}\,dt
           +\left(\bar\eta+\bar\eta^2\beta\right)\alpha^{-1}\frac{d\lambda}{dx^i}\,dx^i,\\
  \widetilde{\Psi}^{II}_0(t)\,dt +\widetilde{\Psi}^{II}_J(t)H^J_{~i}(x^j)\,dx^i &=
  \left( \alpha\dot{\beta}\right)\,dt
           +\left(-\alpha\beta\right)\frac{d\lambda}{dx^i}\,dx^i,
\end{align}
\end{subequations}
where it becomes immediately obvious, that
\begin{equation}
\mathcal{H}^J_{~i}(x^j) \propto \frac{d\lambda}{dx^i}.
\end{equation}
Equivalently, expressed as a one-form, this means
\begin{equation}
d\lambda = \lambda_J \mathcal{H}^J, \label{dlambda}
\end{equation}
where $\lambda_J$ are three complex valued constants. Equation \eqref{dlambda} is integrable if
\begin{equation}
d^2\lambda = \lambda_J \mathcal{C}^J = 0.
\end{equation}
Therefore, the integrability of equation \eqref{dlambda} depends on the Bianchi classification.  For example, for Bianchi type I, $\mathcal{C}^J \equiv 0$ and therefore $\lambda_J$ can be three arbitrary complex valued constants \cite{Coley:2023ibm}.  While for Bianchi types $VIII$ and $IX$, the $\mathcal{C}^J\not = 0$ and therefore equation \eqref{dlambda} is integrable only if $\lambda_J=0$ for $J\in\{1,2,3\}$. For Bianchi type II, $\mathcal{C}^1\not=0$ so $\lambda_1=0$.  For all other Bianchi types, $\mathcal{C}^2\not=0$ and $\mathcal{C}^3\not = 0$ so $\lambda_2=\lambda_3=0$. Knowing the integrability conditions allows one to solve for $\lambda(x^j)$ for each Bianchi type.  Solving equation \eqref{dlambda} for each Bianchi type indicates that
\begin{equation}
\lambda(x^j) = \lambda_1 x + \lambda_2 y + \lambda_3 z = \lambda_j x^j
\end{equation}
a linear function of the spatial coordinates, where some $\lambda_j$ are zero depending on the Bianchi type.
Effectively, the Lorentz parameters making up the general form of the spin connection suitable for Bianchi type tele-parallel geometries are
\begin{subequations}\label{solution spin connection}
  \begin{align}
  A(t,x^j)e^{-i\theta(t,x^j)} &= \alpha(t)e^{\lambda_j x^j} ,\\
  \bar{E}(t,x^j)         &= \bar\eta(t)e^{\lambda_j x^j}, \\
  B(t,x^j)             &= \beta(t)e^{-\lambda_j x^j},
  \end{align}
\end{subequations}
where the restrictions on the complex valued constants $\lambda_j$ for each Bianchi type are
\begin{itemize}
\item $I$: no restrictions on $\lambda_1$, $\lambda_2$ or $\lambda_3$;
\item $II$: $\lambda_1=0$;
\item $IV$, $V$, $VI_0$, $VI_h$, $VII_0$, $VII_h$: $\lambda_2=\lambda_3=0$;
\item $VIII$, $IX$: $\lambda_1=\lambda_2=\lambda_3 = 0$.
\end{itemize}
With a general solution for the Lorentz parameter functions, the coordinates of the general spin connection can easily be computed from the following expressions
\begin{subequations}\label{spinconnectioncomponents}
  \begin{align}
  \Theta_0 &= \frac{\dot\alpha}{\alpha} - 2\bar\eta\dot\beta,
                & \Theta_j &= \lambda_j(1+2\bar\eta\beta), \\
   \Psi^{I}_0 &= (\dot{\bar\eta}-\bar\eta^2\dot\beta)\alpha^{-1},
               & \Psi^I_{j}&=\lambda_j(\bar\eta+\bar\eta^2\beta)\alpha^{-1}, \\
   \Psi^{II}_0 &= \alpha\dot\beta,
             & \Psi^{II}_j &= -\lambda_j(\alpha\beta).
  \end{align}
\end{subequations}

\subsection{Antisymmetric Field Equations for Orthonormal Bianchi Co-frames}

The most general orthonormal co-frame and spin connection having a simply transitive $G_3$ group of affine symmetries acting on spatial hyper-surfaces is given by the geometry
\begin{equation}
\left\{\gortho_{ab},\hortho^a,\Wortho^a_{~b}\right\}
\end{equation}
where $\hortho^a$ is given by equation \eqref{initial-frame} and $\Wortho^a_{~b}$ is computed from equation \eqref{gen_spin_ortho} with coordinates given by \eqref{spinconnectioncomponents}. From here on we will assume we are working in an orthonormal frame and drop the over-circle on such quantities so that they are not confused with time derivatives which will be indicated with an over-dot.
%

The torsion two form is
\begin{equation}
T^a=\left[\omega^{\hat{0}}_{~I}\wedge h^I, \bdot{M}{}^I_{~J}\,dt \wedge \mathcal{H}^J + M^I_{~J}\,\mathcal{C}^J+\omega^I_{~\hat{0}}\wedge dt + \omega^I_{~J}\wedge h^J\right],\label{torsion_calc}
\end{equation}
where we employ the over-dot to denote a time derivative.

Given the simply transitive $G_3$ group of symmetries, the torsion scalar can only be a function of $t$, and assuming that we only have scalar matter, the antisymmetric part of the [FIRST] field equation reduces to
 \begin{equation}
-f''(T)\,\bdot{T}h^{\hat{0}}\wedge {}^*S_{[a}\wedge h_{b]}=0.
\end{equation}
The possible solutions of the above equation divide into three cases:
\begin{itemize}
\item[Case 1:] the torsion scalar is a constant, $T=T_0$,
\item[Case 2:] or $f''(T)=0$ in which case $f(T)=f_1T+f_0$, where $f_1$ and $f_0$ are constants,
\item[Case 3:] or $h^{\hat{0}}\wedge {}^*S_{[a}\wedge h_{b]}=0$.
\end{itemize}

In the first two cases the symmetric part of the [FIRST] field equation reduces
\begin{equation}
-\frac{1}{2}\RLC^{cd}\wedge \eta_{acd}-\frac{1}{2}\tilde{\Lambda}\eta_a = \tilde{\kappa} \Sigma_a,
\end{equation}
where in Case 1,
\begin{equation}
\tilde{\Lambda} = T_0-\frac{f(T_0)}{f'(T_0)}\qquad \mbox{and} \qquad \tilde{\kappa}=\frac{\kappa}{f'(T_0)},
\end{equation}
and for Case 2,
\begin{equation}
\tilde{\Lambda} = -\frac{f_0}{f_1}\qquad \mbox{and} \qquad \tilde{\kappa}=\frac{\kappa}{f_1}.
\end{equation}
In Case 1 and Case 2 the resulting theory is equivalent to GR (or TEGR) with a cosmological constant and a re-scaled gravitational coupling parameter.

In Case 3, where the torsion scalar is not a constant and $f''(T)\not = 0$, we can then split the antisymmetric part of the [FIRST] field equation into the electric ($\hat{0}I$) and the magnetic ($IJ$) components, which reduce to the equivalent of
\begin{subequations}
\begin{eqnarray}
h^{\hat{0}}\wedge {}^*S_{[\hat{0}}\wedge h_{I]}=0 \rightarrow K^J_{~I}\wedge \eta_{J}=0 &\rightarrow& K^J_{~IJ} = 0, \\
h^{\hat{0}}\wedge {}^*S_{[I}\wedge h_{J]}=0 \rightarrow K^{\hat{0}}_{~[I}\wedge \eta_{J]}=0 &\rightarrow& K^{\hat{0}}_{~[IJ]}=0,
\end{eqnarray}
\end{subequations}
in terms of the contorsion. Equivalently the above impose the following constraints on the torsion
\begin{subequations}
\begin{eqnarray}
T^J\wedge \eta_{JI}=0 &\rightarrow& T^{J}_{~IJ}=0\label{asymm_T1},\\
T^{\hat{0}}\wedge \eta_{JI}=0 &\rightarrow& T^{\hat{0}}_{~IJ}=0.\label{asymm_T2}
\end{eqnarray}
\end{subequations}
Substituting \eqref{torsion_calc} in \eqref{asymm_T1} and \eqref{asymm_T2} we find the following constraints between the functions $M^I_{~J}(t)$, $\omega^a_{~b\hat{0}}(t)$, $\omega^a_{~bI}(t)$, the spatial geometry determined by $\mathcal{H}^I(x^i)$, and the constant two forms $\mathcal{C}^J$.
\begin{subequations}
\begin{eqnarray}
M^I_{~K}(t)\,\mathcal{C}^K\wedge\eta_{IJ} + \omega^I_{~J}\wedge \eta_I&=&0,\\
\omega^{\hat{0}}_{~[I}\wedge\eta_{J]}&=&0.
\end{eqnarray}
\end{subequations}
Therefore, the antisymmetric part of the field equations yield six algebraic equations for the spin connection coefficients
\begin{subequations}\label{antisymFE}
\begin{eqnarray}
\omega^K_{~IK}(t)&=&\mathcal{C}^K_{~JK} (M^{-1})^J_{~I}(t), \label{anti1}\\
\omega^{\hat{0}}_{~[IJ]}(t) &=&0, \label{anti2}
\end{eqnarray}
\end{subequations}
in terms of the frame functions $M^I_{~J}(t)$ and the structure constants $\mathcal{C}^I_{~JK}$ defining the Lie algebra for the symmetry group. We will now attempt to find solutions to equations \eqref{antisymFE} for each Bianchi type when the torsion scalar is not a constant and $f''(T)\not = 0$, i.e, Case 3.

\subsection{Bianchi Type A}

If we have a Bianchi type A affine geometry [types $(I,II,VI_0,VII_0,VIII,IX)$] in which $\mathcal{C}^K_{~JK}=0$ (assuming the torsion scalar is not a constant and $f''(T)\not = 0$), then a simple solution to equations \eqref{antisymFE} which also respects the zero curvature and zero nonmetricity constraints is $\omega^a_{~bc}(t)=0$. This is easily accomplished by selecting the Lorentz parameter functions $\alpha(t)=1$, $\eta(t)=0$, $\beta(t)=0$ and the constants $\lambda_1=\lambda_2=\lambda_3=0$  in equation \eqref{spinconnectioncomponents}.  Therefore for a Bianchi type A affine geometry we always have the freedom to choose a proper orthonormal frame, and such a proper orthonormal frame is consistent with the antisymmetric part of the $f(T)$ tele-parallel gravity field equations. In general this proper co-frame will depend on the 9 arbitrary functions of $t$, $M^I_{~J}(t)$ such that the matrix is non-singular.

Alternatively, a different general solution could be constructed simply by letting $\lambda_1=\lambda_2=\lambda_3=0$, in the solution for the Lorentz parameter functions in equation \eqref{solution spin connection}.  In this case, the Lorentz transformation are functions of time only.  Therefore, by applying the corresponding inverse Lorentz transformation we can simply transform the spin connection to zero, resulting in a proper frame, which is equivalent to the case in which we simply set the spin connection to zero as we did initially.  Therefore without loss of generality, for Bianchi type A geometries, the proper frame with the 9 functions $M^I_{~J}(t)$ is a general case.

We do note that other special solutions are possible for Bianchi type A affine geometries.  For example, for the Bianchi type I geometries studied in \cite{Coley:2023ibm}, it was shown that special solutions for the Lorentz transformation exist which depend on the spatial coordinates.  That is, some of the complex-valued constants $\lambda_J$ are not zero.   These special solutions also provide a consistent co-frame/spin connection pair that satisfies the antisymmetric part of the $f(T)$ tele-parallel gravity field equations.

\subsubsection{Bianchi type A Field Equations}
To illustrate the situation further for Bianchi type A geometries, we compute the various quantities and field equations for a class of Bianchi type A geometries. The simplest symmetry aligned proper orthonormal co-frame for the Bianchi type A geometries results from assuming that the matrix $M^I_{~J}(t)$ is diagonal. The co-frame basis is
\begin{equation}
\horthop^a=[dt,A(t)\mathcal{H}^1(x^i),B(t)\mathcal{H}^2(x^i),C(t)\mathcal{H}^3(x^i)]. \label{coframe_bianchiA}
\end{equation}
where the $\mathcal{H}^I(x^i)$ are the canonical co-frame basis in the canonical coordinates $(x^i)$ for each Bianchi type A geometry.
The torsion two-form is
\begin{equation}
T^a=[0,\bdot{A} dt \wedge \mathcal{H}^1 + Ad\mathcal{H}^1, \bdot{B} dt \wedge \mathcal{H}^2 + Bd\mathcal{H}^2,\bdot{C} dt \wedge \mathcal{H}^3 + Cd\mathcal{H}^3].
\end{equation}
The torsion scalar is
\begin{eqnarray}
T &=& 2\left(\frac{\bdot{A}\bdot{B}}{AB}+\frac{\bdot{A}\bdot{C}}{AC}+\frac{\bdot{B}\bdot{C}}{BC}\right)\nonumber\\
&&-\frac{1}{A^2}\left(\mathcal{C}^2_{~31}\mathcal{C}^{31}_{~~2}\right)
-\frac{1}{B^2}\left(\mathcal{C}^1_{~32}\mathcal{C}^{32}_{~~1}\right)
-\frac{1}{C^2}\left(\mathcal{C}^1_{~23}\mathcal{C}^{23}_{~~1}\right)\nonumber\\
&&+\frac{A^2}{B^2C^2}\left(\mathcal{C}^1_{~23}\mathcal{C}_1^{~23}\right)
+\frac{B^2}{A^2C^2}\left(\mathcal{C}^2_{~31}\mathcal{C}_2^{~31}\right)
+\frac{C^2}{A^2B^2}\left(\mathcal{C}^3_{~12}\mathcal{C}_3^{~12}\right),
\end{eqnarray}
where the $\mathcal{C}^I_{~JK}$ are the structure constants for each Bianchi type A which take on values $0,\pm 1$. The torsion scalar can be decomposed into two parts, one that contains time derivatives of the frame functions $A,B,C$, and one that contains the structure constants. For ease of notation we define $T_c$ to be part of $T$ containing the structure constants,
\begin{equation}
T=2\left(\frac{\bdot{A}\bdot{B}}{AB}+\frac{\bdot{A}\bdot{C}}{AC}+\frac{\bdot{B}\bdot{C}}{BC}\right)+T_c,
\label{T_bianchiA}
\end{equation}
where the expressions for $T_c$ are given in Table \ref{Table1}.

\begin{table}
  \centering
  {\renewcommand{\arraystretch}{2}
  \begin{tabular}{|l|c|}
    \hline
    Bianchi type & $T_c$  \\
    \hline
    $I$ & 0  \\
    $II$ & $\displaystyle\frac{A^2}{2B^2C^2}$  \\
    $VI_0$ & $\displaystyle\frac{(B^2+C^2)^2}{2A^2B^2C^2}$ \\
    $VII_0$ & $\displaystyle\frac{(B^2-C^2)^2}{2A^2B^2C^2}$ \\
    $VIII$ & $\displaystyle\frac{(A^2+(B-C)^2)(A^2+(B+C)^2)}{2A^2B^2C^2}$ \\
    $IX$ & $\displaystyle\frac{(A^2-(B-C)^2)(A^2-(B+C)^2)}{2A^2B^2C^2}$ \\[1 em]
    \hline
  \end{tabular}
  }
  \caption{The structure constant part $T_c$ of the Torsion Scalar $T$ for each diagonal Bianchi type A affine symmetry.}\label{Table1}
\end{table}

For completeness we express the $f(T)$ tele-parallel gravity field equations for the symmetry aligned, diagonal, proper, orthonormal frame for the Bianchi type A geometries described by the co-frame \eqref{coframe_bianchiA}. Assuming a co-moving perfect fluid source the symmetric part of the field equations \eqref{sym_first1} are
\begin{subequations}\label{FE for Bianchi A}
  \begin{align}
 f'(T)\left( \frac{\bdot A}{A}\frac{\bdot B}{B} + \frac{\bdot A}{A}\frac{\bdot C}{C}+\frac{\bdot B}{B}\frac{\bdot C}{C}-\frac{1}{2}T_c\right) - \frac{1}{2}\left(f(T)-Tf'(T)\right) &= \kappa \rho \\
  f'(T)\left( -\frac{\bddot B}{B} - \frac{\bddot C}{C} -\frac{\bdot B}{B}\frac{\bdot C}{C}-\frac{1}{2}T_c+s_1\right) + \frac{1}{2}\left(f(T)-Tf'(T)\right) -f''(T)\bdot{T}\left(\frac{\bdot A}{A}+\frac{\bdot B}{B}\right)&= \kappa p \\
  f'(T)\left( -\frac{\bddot A}{A} - \frac{\bddot C}{C} -\frac{\bdot A}{A}\frac{\bdot C}{C}-\frac{1}{2}T_c+s_2\right) + \frac{1}{2}\left(f(T)-Tf'(T)\right) -f''(T)\bdot{T}\left(\frac{\bdot A}{A}+\frac{\bdot C}{C}\right)&= \kappa p \\
  f'(T)\left( -\frac{\bddot A}{A} - \frac{\bddot B}{B} -\frac{\bdot A}{A}\frac{\bdot B}{B}-\frac{1}{2}T_c+s_3\right) + \frac{1}{2}\left(f(T)-Tf'(T)\right) -f''(T)\bdot{T}\left(\frac{\bdot B}{B}+\frac{\bdot C}{C}\right)&= \kappa p
  \end{align}
\end{subequations}
The values of $s_i$ appearing in the field equations depend on the specific Bianchi type: expressions for each can be found in Table \ref{TableA2}. We reiterate, the antisymmetric part of field equations are identically satisfied for Bianchi type A geometries with a proper frame.

\begin{table}
\centering
  {\renewcommand{\arraystretch}{2}
\begin{tabular}{|l|c|c|c|}
  \hline
  Bianchi Type & $s_1$ & $s_2$ & $s_3$  \\
  \hline
  $I$ & 0 & 0 & 0 \\
  $II$ & $\displaystyle\frac{A^2}{B^2 C^2}$ & 0 & 0 \\
  $VI_0$ & 0 & $\displaystyle\frac{B^2+C^2}{A^2 C^2}$ & $\displaystyle\frac{B^2+C^2}{A^2 B^2}$ \\
  $VII_0$ & 0 & $\displaystyle\frac{B^2-C^2}{A^2 C^2}$ & $\displaystyle\frac{-B^2+C^2}{A^2 B^2}$\\
  $VIII$ & $\displaystyle\frac{A^2+B^2+C^2}{B^2 C^2}$  & $\displaystyle\frac{A^2+B^2-C^2}{A^2 C^2}$ & $\displaystyle\frac{A^2-B^2+C^2}{A^2 B^2}$\\
  $IX$ & $\displaystyle\frac{A^2-B^2-C^2}{B^2 C^2}$  & $\displaystyle\frac{-A^2+B^2-C^2}{A^2 C^2}$ & $\displaystyle\frac{-A^2-B^2+C^2}{A^2 B^2}$ \\[1 em]
  \hline
\end{tabular}
\caption{The field equations for $f(T)$ tele-parallel gravity for Bianchi type A affine frame symmetries depend on three functions $s_i$ constructed from the structure constants of the associated Lie algebra and the frame functions $A$, $B$, and $C$.}}\label{TableA2}
\end{table}

\subsection{Bianchi Type B}
If we have a Bianchi type B affine geometry [types $(IV,V,VI_h,VII_h)$] in which $\mathcal{C}^K_{~JK}\not=0$ (assuming the torsion scalar is not a constant and $f''(T)\not = 0$), then it is necessary that some of the $\omega^a_{~bc}(t)\not = 0$.   We will take a closer look at equations \eqref{antisymFE}.  For $I=1,2,3$ we obtain the following three expressions for equation \eqref{anti1}
\begin{subequations}\label{antisym1}
\begin{align}
-Re(\Psi^I-\Psi^{II})_\mu e_{\hat{2}}^{~\mu} + Im(\Psi^I + \Psi^{II})_\mu e_{\hat{3}}^{~\mu} &= \mathcal{C}^K_{~JK}(M^{-1})^J_{~\hat{1}},\\
 Re(\Psi^I-\Psi^{II})_\mu e_{\hat{1}}^{~\mu} - Im(\Theta)_\mu e_{\hat{3}}^{~\mu} &= \mathcal{C}^K_{~JK}(M^{-1})^J_{~\hat{2}},\\
 -Im(\Psi^I+\Psi^{II})_\mu e_{\hat{1}}^{~\mu} + Im(\Theta)_\mu e_{\hat{2}}^{~\mu} &= \mathcal{C}^K_{~JK}(M^{-1})^J_{~\hat{3}},
 \end{align}
\end{subequations}
and from equation \eqref{anti2} we find
\begin{subequations}\label{antisym2}
\begin{align}
 Re(\Theta)_\mu e_{\hat{2}}^{~\mu} - Re(\Psi^I + \Psi^{II})_\mu e_{\hat{1}}^{~\mu} &= 0,\\
 Re(\Theta)_\mu e_{\hat{3}}^{~\mu} + Im(\Psi^I - \Psi^{II}) e_{\hat{1}}^{~\mu} &= 0,\\
 Re(\Psi^I+\Psi^{II})_\mu e_{\hat{3}}^{~\mu} + Im(\Psi^I - \Psi^{II}) e_{\hat{2}}^{~\mu} &= 0.
 \end{align}
\end{subequations}
Given that $e_I$ is a vector tangent to the spatial hyper-surface, $e_I^{~0}=0$. We can combine equations \eqref{antisym1} and \eqref{antisym2} to obtain three equivalent complex valued equations
\begin{subequations}\label{complex antisym}
  \begin{align}
  (\Psi^I+\Psi^{II})_j e_{\hat{3}}^{~j} -i (\Psi^I - \Psi^{II})_j e_{\hat{2}}^{~j} &= iC^K_{~JK}(M^{-1})^J_{~\hat{1}},\\
 +i (\Psi^I - \Psi^{II})_j e_{\hat{1}}^{~j} -(\Theta)_j e_{\hat{3}}^{~j}  &= iC^K_{~JK}(M^{-1})^J_{~\hat{2}},\\
- (\Psi^I+\Psi^{II})_j e_{\hat{1}}^{~j} +  (\Theta)_j e_{\hat{2}}^{~j} &= iC^K_{~JK}(M^{-1})^J_{~\hat{3}}.
 \end{align}
\end{subequations}
For Bianchi type B geometries,
\begin{equation}
\mathcal{C}^K_{~JK}=2c\delta^1_{J}, \qquad\text{where\ } c = \left\{ \begin{array}{cl} 1, & \text{Bianchi\ } IV, V \\
\sqrt{|h|},\ \  & \text{Bianchi\ } VI_h, VII_h
\end{array}\right.
\end{equation}
Further in all four Bianchi type B geometries, we have from solving for the spin connection in Section \ref{sec:spin} that $\lambda_2=\lambda_3=0$, which implies $\Theta_y=\Theta_z=0$, $\Psi^I_y=\Psi^I_z=0$ and $\Psi^{II}_y=\Psi^{II}_z=0$ so equations \eqref{complex antisym} become
\begin{subequations}\label{complex antisym2}
  \begin{align}
  (\Psi^I_x+\Psi^{II}_x) e_{\hat{3}}^{~x} -i (\Psi^I_x - \Psi^{II}_x) e_{\hat{2}}^{~x} &= 2ic(M^{-1})^{\hat{1}}_{~\hat{1}},\\
  +i (\Psi^I_x - \Psi^{II}_x) e_{\hat{1}}^{~x} -(\Theta_x) e_{\hat{3}}^{~x}  &= 2ic(M^{-1})^{\hat{1}}_{~\hat{2}},\\
  - (\Psi^I_x+\Psi^{II}_x) e_{\hat{1}}^{~x} +(\Theta_x) e_{\hat{2}}^{~x} &= 2ic(M^{-1})^{\hat{1}}_{~\hat{3}}.
 \end{align}
\end{subequations}
The above system can be considered a linear system for $\Psi^I_x+\Psi^{II}_x$, $\Psi^I_x-\Psi^{II}_x$ and  $\Theta_x$.  Indeed we can express it in matrix format as
\begin{equation} \label{matrixeqn}
\begin{bmatrix}
  e_{\hat{3}}^{~x} &  - ie_{\hat{2}}^{~x} & 0 \\
  0                 &  ie_{\hat{1}}^{~x}  & -e_{\hat{3}}^{~x}  \\
  -e_{\hat{1}}^{~x} & 0                   &  e_{\hat{2}}^{~x}
\end{bmatrix}
\begin{bmatrix}
  \Psi^I_x+\Psi^{II}_x \\
  \Psi^I_x-\Psi^{II}_x \\
  \Theta_x
\end{bmatrix}
=2ic
\begin{bmatrix}
  (M^{-1})^{\hat{1}}_{~\hat{1}} \\
  (M^{-1})^{\hat{1}}_{~\hat{2}} \\
  (M^{-1})^{\hat{1}}_{~\hat{3}}
\end{bmatrix}
\end{equation}
where we note
\begin{equation}\label{HIX}
e_{I}^{~x}=(M^{-1})^J_{~I}\mathcal{E}_J^{~x}=(M^{-1})^{\hat{1}}_{~I},
\end{equation}
since  $\mathcal{E}_{\hat{1}}^{~x} =1$, $\mathcal{E}_{\hat{2}}^{~x} =0$ and $\mathcal{E}_{\hat{3}}^{~x} =0$   for Bianchi type B geometries.

The determinant of the coefficient matrix in equation \eqref{matrixeqn} is zero, and therefore there is no solution in general for Bianchi type B tele-parallel geometries.  This means that in general, Bianchi type B tele-parallel geometries (with $T\not = \text{const}$ and $f''(T)\not = 0$) are inconsistent with the antisymmetric part of the $f(T)$ tele-parallel field equations, that is; no general solutions exist.

Of course, special degenerate solutions may exist if the rank of the matrix is equal to the rank of the augmented matrix. This can occur if the degeneracy condition
\begin{equation}\label{specsol}
(M^{-1})^{\hat{1}}_{~\hat{1}}e_{\hat{1}}^{~x}+(M^{-1})^{\hat{1}}_{~\hat{2}}e_{\hat{2}}^{~x}
    +(M^{-1})^{\hat{1}}_{~\hat{3}}e_{\hat{3}}^{~x}=0
\end{equation}
is met. When equation \eqref{HIX} is substituted into the degeneracy condition \eqref{specsol} we obtain
\begin{equation}
\left((M^{-1})^{\hat{1}}_{~\hat{1}}\right)^2  + \left((M^{-1})^{\hat{1}}_{~\hat{2}}\right)^2  + \left((M^{-1})^{\hat{1}}_{~\hat{3}}\right)^2  =0,
\end{equation}
solutions of which imply $(M^{-1})^{\hat{1}}_{~\hat{1}}=(M^{-1})^{\hat{1}}_{~\hat{2}}=(M^{-1})^{\hat{1}}_{~\hat{3}}=0$. This yields a matrix $(M^{-1})^I_{~J}$ with a column of zeros which is not possible since the matrix $M^I_{~J}\in GL(3)$ is non-singular. Therefore, equation \eqref{specsol} is never true.   The Bianchi type B tele-parallel geometries are inconsistent with the $f(T)$ tele-parallel gravity unless $f(T)$ is a linear function of $T$ or $T$ is constant; that is, the theory is TEGR (or GR) with a cosmological constant and a re-scaled coupling constant.


\section{Concluding remarks}

Employing the Metric Affine Gauge framework the field equations for $f(T)$ tele-parallel gravity were developed in a fully covariant manner through the use of Lagrange multipliers. Further, with the assumption of a scalar matter source, it was shown that the covariant field equations can be split into a symmetric part which has a matter source, and an antisymmetric part with no source.  The anti-symmetric part of the $f(T)$ tele-parallel gravity field equations can place severe constraints on the function $f(T)$ if the geometrical framework is not properly initialized.  For example, if the geometry is Minkowski expressed in spherical coordinates, then using a diagonal co-frame and a trivial spin connection implies that the function $f(T)$ must be linear in $T$.  Essentially, the initial setup for this problem is incorrect.

The goal is to develop co-frame/spin connection pairs suitable for Bianchi affine geometries and use these co-frame/spin connection pairs in the covariant field equations for $f(T)$ tele-parallel gravity. Assuming the co-frame and spin connection are both invariant under a simply transitive $G_3$ group of motions an orthonormal co-frame basis is constructed.  Then generalizing the results in \cite{Coley:2023ibm}, a spin connection suitable for any Bianchi affine geometry is computed: this spin connection being a function of three complex valued functions of time and three arbitrary complex valued constants.  Depending on the Bianchi type, some of these constants are zero.

With the geometrical framework developed for each Bianchi type, the antisymmetric part of the $f(T)$ tele-parallel gravity field equations is calculated.  For Bianchi type A affine geometries with a perfect fluid source moving orthogonally to the spatial hypersurfaces, the proper frame is consistent with the anti-symmetric part of the field equations; that is, the antisymmetric part of the field equations are identically satisfied.  A consistent proper co-frame always exists in general for Bianchi type A affine geometries; that is, a so-called ``good'' co-frame exists.  Assuming a diagonal proper co-frame, the symmetric part of the $f(T)$ tele-parallel gravity field equations with a perfect fluid are expressed for each Bianchi type A geometry.

For Bianchi type B affine geometries we find a different result. With the assumption of a perfect fluid source moving orthogonally to the spatial hypersurfaces the antisymmetric part of the $f(T)$ tele-parallel gravity field equations can be expressed as a non-homogeneous matrix problem. However, it was shown that no solutions exist unless $f(T)=f_1T+f_0$ or $T$ is a constant. The Bianchi type B tele-parallel geometries are inconsistent with the $f(T)$ tele-parallel gravity unless the theory is TEGR (or GR) with a cosmological constant and a re-scaled coupling constant.  An analogous inconsistency regarding Bianchi type B geometries is also described in \cite{Maccallum:1972er} where it is shown that the general action principles are not valid for Bianchi type B spatially-homogeneous geometries if the symmetry assumptions are imposed before the variations are made.  Here, we show that the antisymmetric part of the $f(T)$ tele-parallel gravity field equations can never be satisfied for general Bianchi type B affine geometries containing a perfect fluid moving orthogonally to the spatial hypersurfaces. This result of course does depends on the nature of the matter. If the matter has spin, then there exists a matter source in the antisymmetric part of the $f(T)$ tele-parallel gravity field equations.

One important question in mathematical cosmology is that of initial conditions.  Most often, solutions to the field equations assume from the onset that the geometry is both spatially homogeneous and isotropic on the largest of scales, but this is an extremely restrictive assumption.  Presumably the initial conditions were such that the universe was inhomogeneous and evolved toward spatial homogeneity and isotropy.  Therefore building inhomogeneous and/or anisotropic models is of import.  Here, we build a framework that will allow one to construct large classes of Bianchi type A anisotropic models which can then be subsequently analyzed and tested against current observations.

Obviously there is more work to do.  For example, one remaining spatially homogenous but anisotropic geometry to be constructed in tele-parallel gravity is that of a $G_4$ acting on the 3 dimensional spatial hyper-surface (tele-parallel Kantowski-Sachs).  This geometry has an isotropy subgroup, and therefore the techniques developed here and in \cite{McNutt:2023nxm} will need to be employed.



\begin{acknowledgments}
AAC and RvdH are supported by the Natural Sciences and Engineering Research Council of Canada. RvdH is supported by the Dr. W.F. James Chair of Studies in the Pure and Applied Sciences at St.F.X.  DDM is supported by the Norwegian Financial Mechanism 2014-2021 (project registration number 2019/34/H/ST1/00636).
\end{acknowledgments}


\bibliographystyle{JHEP}
\bibliography{../BIBFILES/Tele-Parallel-Reference-file}

\providecommand{\href}[2]{#2}\begingroup\raggedright\begin{thebibliography}{10}

\bibitem{Will:2014kxa}
C.M.~Will, \emph{{The Confrontation between General Relativity and
  Experiment}}, \href{https://doi.org/10.12942/lrr-2014-4}{\emph{Living Rev.
  Rel.} {\bfseries 17} (2014) 4}
  [\href{https://arxiv.org/abs/1403.7377}{{\ttfamily 1403.7377}}].

\bibitem{WMAP:2012nax}
{\scshape WMAP} collaboration, \emph{{Nine-Year Wilkinson Microwave Anisotropy
  Probe (WMAP) Observations: Cosmological Parameter Results}},
  \href{https://doi.org/10.1088/0067-0049/208/2/19}{\emph{Astrophys. J. Suppl.}
  {\bfseries 208} (2013) 19} [\href{https://arxiv.org/abs/1212.5226}{{\ttfamily
  1212.5226}}].

\bibitem{Planck:2018vyg}
{\scshape Planck} collaboration, \emph{{Planck 2018 results. VI. Cosmological
  parameters}},
  \href{https://doi.org/10.1051/0004-6361/201833910}{\emph{Astron. Astrophys.}
  {\bfseries 641} (2020) A6}
  [\href{https://arxiv.org/abs/1807.06209}{{\ttfamily 1807.06209}}].

\bibitem{Nojiri:2010wj}
S.~Nojiri and S.D.~Odintsov, \emph{{Unified cosmic history in modified gravity:
  from F(R) theory to Lorentz non-invariant models}},
  \href{https://doi.org/10.1016/j.physrep.2011.04.001}{\emph{Phys. Rept.}
  {\bfseries 505} (2011) 59} [\href{https://arxiv.org/abs/1011.0544}{{\ttfamily
  1011.0544}}].

\bibitem{Nojiri:2005jg}
S.~Nojiri and S.D.~Odintsov, \emph{{Modified Gauss-Bonnet theory as
  gravitational alternative for dark energy}},
  \href{https://doi.org/10.1016/j.physletb.2005.10.010}{\emph{Phys. Lett.}
  {\bfseries B631} (2005) 1}
  [\href{https://arxiv.org/abs/hep-th/0508049}{{\ttfamily hep-th/0508049}}].

\bibitem{Nojiri_Odintsov2006}
S.~Nojiri and S.D.~Odintsov, \emph{Introduction to modified gravity and
  gravitational alternative for dark energy},
  \href{https://doi.org/10.1142/S0219887807001928}{\emph{International Journal
  of Geometric Methods in Modern Physics} {\bfseries 04} (2007) 115}
  [\href{https://arxiv.org/abs/0601213}{{\ttfamily 0601213}}].

\bibitem{Capozziello_DeLaurentis_2011}
S.~Capozziello and M.~De~Laurentis, \emph{Extended theories of gravity},
  \href{https://doi.org/http://dx.doi.org/10.1016/j.physrep.2011.09.003}{\emph{Physics
  Reports} {\bfseries 509} (2011) 167 }
  [\href{https://arxiv.org/abs/gr-qc/1108.6266}{{\ttfamily gr-qc/1108.6266}}].

\bibitem{Clifton:2011jh}
T.~Clifton, P.G.~Ferreira, A.~Padilla and C.~Skordis, \emph{{Modified Gravity
  and Cosmology}},
  \href{https://doi.org/10.1016/j.physrep.2012.01.001}{\emph{Physics Reports}
  {\bfseries 513} (2012) 1} [\href{https://arxiv.org/abs/1106.2476}{{\ttfamily
  1106.2476}}].

\bibitem{CANTATA:2021ktz}
{\scshape CANTATA} collaboration, \emph{{Modified Gravity and Cosmology: An
  Update by the CANTATA Network}},
  \href{https://arxiv.org/abs/2105.12582}{{\ttfamily 2105.12582}}.

\bibitem{Aldrovandi_Pereira2013}
R.~Aldrovandi and J.G.~Pereira, \emph{{Teleparallel Gravity}}, vol.~173 of
  \emph{Fundamental Theories of Physics}, Springer, Dordrecht (2013),
  \href{https://doi.org/10.1007/978-94-007-5143-9}{10.1007/978-94-007-5143-9}.

\bibitem{Krssak:2018ywd}
M.~Krssak, R.J.~van~den Hoogen, J.G.~Pereira, C.G.~Boehmer and A.A.~Coley,
  \emph{{Teleparallel Theories of Gravity: Illuminating a Fully Invariant
  Approach}}, \href{https://doi.org/10.1088/1361-6382/ab2e1f}{\emph{Classical
  and Quantum Gravity} {\bfseries 36} (2019) 183001}
  [\href{https://arxiv.org/abs/1810.12932}{{\ttfamily 1810.12932}}].

\bibitem{Krssak_Saridakis2015}
M.~Kr\v{s}\v{s}\'{a}k and E.N.~Saridakis, \emph{{The covariant formulation of
  f(T) gravity}},
  \href{https://doi.org/10.1088/0264-9381/33/11/115009}{\emph{Classical and
  Quantum Gravity} {\bfseries 33} (2016) 115009}
  [\href{https://arxiv.org/abs/1510.08432}{{\ttfamily 1510.08432}}].

\bibitem{Obukhov_Pereira2003}
Y.N.~Obukhov and J.G.~Pereira, \emph{{Metric affine approach to teleparallel
  gravity}}, \href{https://doi.org/10.1103/PhysRevD.67.044016}{\emph{Physical
  Review} {\bfseries D67} (2003) 044016}
  [\href{https://arxiv.org/abs/gr-qc/0212080}{{\ttfamily gr-qc/0212080}}].

\bibitem{Maluf2013}
J.W.~Maluf, \emph{{The teleparallel equivalent of general relativity}},
  \href{https://doi.org/10.1002/andp.201200272}{\emph{Annalen Phys.} {\bfseries
  525} (2013) 339} [\href{https://arxiv.org/abs/1303.3897}{{\ttfamily
  1303.3897}}].

\bibitem{MuellerHoissen_Nitsch1983}
F.~Mueller-Hoissen and J.~Nitsch, \emph{{TELEPARALLELISM - A VIABLE THEORY OF
  GRAVITY?}}, \href{https://doi.org/10.1103/PhysRevD.28.718}{\emph{Phys. Rev.}
  {\bfseries D28} (1983) 718}.

\bibitem{Hayashi:1979qx}
K.~Hayashi and T.~Shirafuji, \emph{{New General Relativity}},
  \href{https://doi.org/10.1103/PhysRevD.19.3524}{\emph{Phys. Rev. D}
  {\bfseries 19} (1979) 3524}.

\bibitem{Ferraro:2006jd}
R.~Ferraro and F.~Fiorini, \emph{{Modified teleparallel gravity: Inflation
  without inflaton}},
  \href{https://doi.org/10.1103/PhysRevD.75.084031}{\emph{Physical Review}
  {\bfseries D75} (2007) 084031}
  [\href{https://arxiv.org/abs/gr-qc/0610067}{{\ttfamily gr-qc/0610067}}].

\bibitem{Ferraro:2008ey}
R.~Ferraro and F.~Fiorini, \emph{{On Born-Infeld Gravity in Weitzenbock
  spacetime}}, \href{https://doi.org/10.1103/PhysRevD.78.124019}{\emph{Phys.
  Rev.} {\bfseries D78} (2008) 124019}
  [\href{https://arxiv.org/abs/0812.1981}{{\ttfamily 0812.1981}}].

\bibitem{Linder:2010py}
E.V.~Linder, \emph{{Einstein's Other Gravity and the Acceleration of the
  Universe}}, \href{https://doi.org/10.1103/PhysRevD.81.127301}{\emph{Phys.
  Rev. D} {\bfseries 81} (2010) 127301}
  [\href{https://arxiv.org/abs/1005.3039}{{\ttfamily 1005.3039}}].

\bibitem{Cai_2015}
Y.-F.~Cai, S.~Capozziello, M.~De~Laurentis and E.N.~Saridakis, \emph{{f(T)
  teleparallel gravity and cosmology}},
  \href{https://doi.org/10.1088/0034-4885/79/10/106901}{\emph{Rept. Prog.
  Phys.} {\bfseries 79} (2016) 106901}
  [\href{https://arxiv.org/abs/1511.07586}{{\ttfamily 1511.07586}}].

\bibitem{Bahamonde:2021gfp}
S.~Bahamonde, K.F.~Dialektopoulos, C.~Escamilla-Rivera, G.~Farrugia, V.~Gakis,
  M.~Hendry et~al., \emph{{Teleparallel gravity: from theory to cosmology}},
  \href{https://doi.org/10.1088/1361-6633/ac9cef}{\emph{Rept. Prog. Phys.}
  {\bfseries 86} (2023) 026901}
  [\href{https://arxiv.org/abs/2106.13793}{{\ttfamily 2106.13793}}].

\bibitem{Hohmann:2019nat}
M.~Hohmann, L.~J\"arv, M.~Kr\v{s}\v{s}\'ak and C.~Pfeifer, \emph{{Modified
  teleparallel theories of gravity in symmetric spacetimes}},
  \href{https://doi.org/10.1103/PhysRevD.100.084002}{\emph{Phys. Rev. D}
  {\bfseries 100} (2019) 084002}
  [\href{https://arxiv.org/abs/1901.05472}{{\ttfamily 1901.05472}}].

\bibitem{Hohmann:2021ast}
M.~Hohmann, \emph{{General covariant symmetric teleparallel cosmology}},
  \href{https://doi.org/10.1103/PhysRevD.104.124077}{\emph{Phys. Rev. D}
  {\bfseries 104} (2021) 124077}
  [\href{https://arxiv.org/abs/2109.01525}{{\ttfamily 2109.01525}}].

\bibitem{DAmbrosio:2021zpm}
F.~D'Ambrosio, S.D.B.~Fell, L.~Heisenberg and S.~Kuhn, \emph{{Black holes in
  f(Q) gravity}},
  \href{https://doi.org/10.1103/PhysRevD.105.024042}{\emph{Phys. Rev. D}
  {\bfseries 105} (2022) 024042}
  [\href{https://arxiv.org/abs/2109.03174}{{\ttfamily 2109.03174}}].

\bibitem{DAmbrosio:2021pnd}
F.~D'Ambrosio, L.~Heisenberg and S.~Kuhn, \emph{{Revisiting cosmologies in
  teleparallelism}},
  \href{https://doi.org/10.1088/1361-6382/ac3f99}{\emph{Class. Quant. Grav.}
  {\bfseries 39} (2022) 025013}
  [\href{https://arxiv.org/abs/2109.04209}{{\ttfamily 2109.04209}}].

\bibitem{Coley:2022qug}
A.A.~Coley, R.J.~van~den Hoogen and D.D.~McNutt, \emph{{Symmetric teleparallel
  geometries}}, \href{https://doi.org/10.1088/1361-6382/ac994a}{\emph{Class.
  Quant. Grav.} {\bfseries 39} (2022) 22LT01}
  [\href{https://arxiv.org/abs/2205.10719}{{\ttfamily 2205.10719}}].

\bibitem{Hehl_McCrea_Mielke_Neeman1995}
F.W.~Hehl, J.D.~McCrea, E.W.~Mielke and Y.~Ne'eman, \emph{{Metric affine gauge
  theory of gravity: Field equations, Noether identities, world spinors, and
  breaking of dilation invariance}},
  \href{https://doi.org/10.1016/0370-1573(94)00111-F}{\emph{Physics Reports}
  {\bfseries 258} (1995) 1}
  [\href{https://arxiv.org/abs/gr-qc/9402012}{{\ttfamily gr-qc/9402012}}].

\bibitem{Rodrigues:2012qua}
M.E.~Rodrigues, M.J.S.~Houndjo, D.~Saez-Gomez and F.~Rahaman,
  \emph{{Anisotropic Universe Models in f(T) Gravity}},
  \href{https://doi.org/10.1103/PhysRevD.86.104059}{\emph{Phys. Rev. D}
  {\bfseries 86} (2012) 104059}
  [\href{https://arxiv.org/abs/1209.4859}{{\ttfamily 1209.4859}}].

\bibitem{Rodrigues:2013iua}
M.E.~Rodrigues, I.G.~Salako, M.J.S.~Houndjo and J.~Tossa, \emph{{Locally
  Rotationally Symmetric Bianchi Type-I cosmological model in $f(T)$ gravity:
  from early to Dark Energy dominated universe}},
  \href{https://doi.org/10.1142/S0218271814500047}{\emph{Int. J. Mod. Phys. D}
  {\bfseries 23} (2014) 1450004}
  [\href{https://arxiv.org/abs/1308.2962}{{\ttfamily 1308.2962}}].

\bibitem{Rodrigues:2014xam}
M.E.~Rodrigues, A.V.~Kpadonou, F.~Rahaman, P.J.~Oliveira and M.J.S.~Houndjo,
  \emph{{Bianchi type-I, type-III and Kantowski-Sachs solutions in $f(T)$
  gravity}}, \href{https://doi.org/10.1007/s10509-015-2358-8}{\emph{Astrophys.
  Space Sci.} {\bfseries 357} (2015) 129}
  [\href{https://arxiv.org/abs/1408.2689}{{\ttfamily 1408.2689}}].

\bibitem{Sharif:2011bi}
M.~Sharif and S.~Rani, \emph{{F(T) Models within Bianchi Type I Universe}},
  \href{https://doi.org/10.1142/S0217732311036127}{\emph{Mod. Phys. Lett. A}
  {\bfseries 26} (2011) 1657}
  [\href{https://arxiv.org/abs/1105.6228}{{\ttfamily 1105.6228}}].

\bibitem{Amir:2015wja}
M.J.~Amir and M.~Yussouf, \emph{{Kantowski-Sachs Universe Models in $f(T)$
  Theory of Gravity}},
  \href{https://doi.org/10.1007/s10773-015-2517-2}{\emph{Int. J. Theor. Phys.}
  {\bfseries 54} (2015) 2798}
  [\href{https://arxiv.org/abs/1502.00777}{{\ttfamily 1502.00777}}].

\bibitem{Fayaz:2014swa}
V.~Fayaz, H.~Hossienkhani, A.~Farmany, M.~Amirabadi and N.~Azimi,
  \emph{{Cosmology of $f(T)$ gravity in a holographic dark energy and
  nonisotropic background}},
  \href{https://doi.org/10.1007/s10509-014-1817-y}{\emph{Astrophys. Space Sci.}
  {\bfseries 351} (2014) 299}.

\bibitem{Fayaz:2015yka}
V.~Fayaz, H.~Hossienkhani, A.~Pasqua, M.~Amirabadi and M.~Ganji, \emph{{f(T)
  theories from holographic dark energy models within Bianchi type I
  universe}}, \href{https://doi.org/10.1140/epjp/i2015-15028-2}{\emph{Eur.
  Phys. J. Plus} {\bfseries 130} (2015) 28}.

\bibitem{Skugoreva:2017vde}
M.A.~Skugoreva and A.V.~Toporensky, \emph{{On Kasner solution in Bianchi I
  $f(T)$ cosmology}},
  \href{https://doi.org/10.1140/epjc/s10052-018-5857-2}{\emph{Eur. Phys. J. C}
  {\bfseries 78} (2018) 377}
  [\href{https://arxiv.org/abs/1711.07069}{{\ttfamily 1711.07069}}].

\bibitem{Skugoreva:2019bwt}
M.A.~Skugoreva and A.V.~Toporensky, \emph{{Anisotropic cosmological dynamics in
  $f(T)$ gravity in the presence of a perfect fluid}},
  \href{https://doi.org/10.1140/epjc/s10052-019-7251-0}{\emph{Eur. Phys. J. C}
  {\bfseries 79} (2019) 813}
  [\href{https://arxiv.org/abs/1907.12538}{{\ttfamily 1907.12538}}].

\bibitem{Tretyakov:2021cgb}
P.V.~Tretyakov, \emph{{Bianchi I cosmological solutions in teleparallel
  gravity}}, \href{https://doi.org/10.1142/S0217732322500468}{\emph{Mod. Phys.
  Lett. A} {\bfseries 37} (2022) 2250046}
  [\href{https://arxiv.org/abs/2109.14457}{{\ttfamily 2109.14457}}].

\bibitem{Aslam:2013coa}
A.~Aslam, M.~Jamil and R.~Myrzakulov, \emph{{Noether gauge symmetry for the
  Bianchi type I model in $f(T)$ gravity}},
  \href{https://doi.org/10.1088/0031-8949/88/02/025003}{\emph{Phys. Scripta}
  {\bfseries 88} (2013) 025003}
  [\href{https://arxiv.org/abs/1308.0325}{{\ttfamily 1308.0325}}].

\bibitem{Paliathanasis:2016vsw}
A.~Paliathanasis, J.D.~Barrow and P.G.L.~Leach, \emph{{Cosmological Solutions
  of $f(T)$ Gravity}},
  \href{https://doi.org/10.1103/PhysRevD.94.023525}{\emph{Phys. Rev. D}
  {\bfseries 94} (2016) 023525}
  [\href{https://arxiv.org/abs/1606.00659}{{\ttfamily 1606.00659}}].

\bibitem{Coley:2023ibm}
A.A.~Coley and R.J.~van~den Hoogen, \emph{{Spatially Homogeneous Teleparallel
  Gravity: Bianchi I}},  \href{https://arxiv.org/abs/2305.12168}{{\ttfamily
  2305.12168}}.

\bibitem{Tamanini_Boehmer2012}
N.~Tamanini and C.G.~B\"{o}hmer, \emph{{Good and bad tetrads in f(T) gravity}},
  \href{https://doi.org/10.1103/PhysRevD.86.044009}{\emph{Phys. Rev.}
  {\bfseries D86} (2012) 044009}
  [\href{https://arxiv.org/abs/1204.4593}{{\ttfamily 1204.4593}}].

\bibitem{Hohmann:2023sto}
M.~Hohmann, \emph{{Spatially homogeneous teleparallel spacetimes with
  four-dimensional groups of motions}},  5, 2023
  [\href{https://arxiv.org/abs/2305.06997}{{\ttfamily 2305.06997}}].

\bibitem{Coley:2019zld}
A.A.~Coley, R.J.~van~den Hoogen and D.D.~McNutt, \emph{{Symmetry and
  Equivalence in Teleparallel Gravity}},
  \href{https://doi.org/10.1063/5.0003252}{\emph{J. Math. Phys.} {\bfseries 61}
  (2020) 072503} [\href{https://arxiv.org/abs/1911.03893}{{\ttfamily
  1911.03893}}].

\bibitem{McNutt:2023nxm}
D.D.~McNutt, A.A.~Coley and R.J.~{van den Hoogen}, \emph{{A frame based
  approach to computing symmetries with non-trivial isotropy groups}},
  \href{https://doi.org/10.1063/5.0134596}{\emph{J. Math. Phys.} {\bfseries 64}
  (2023) 032503} [\href{https://arxiv.org/abs/2302.11493}{{\ttfamily
  2302.11493}}].

\bibitem{Ortin:2004ms}
T.~Ortin, \emph{{Gravity and strings}}, Cambridge Monographs on Mathematical
  Physics, Cambridge Univ. Press (2004),
  \href{https://doi.org/10.1017/CBO9780511616563}{10.1017/CBO9780511616563}.

\bibitem{Ciufolini_Wheeler1995}
I.~Ciufolini and J.~Wheeler, \emph{Gravitation and Inertia}, Princeton series
  in physics, Princeton University Press (1995).

\bibitem{vonderHeyde1975}
P.~von~ver Heyde, \emph{{The Equivalence Principle in the $U_4$ Theory of
  Gravitation}}, {\emph{Lettere Al Nuovo Cimento} {\bfseries 14} (1975) 250}.

\bibitem{Hehl_vonderHeyde_Kerlick_Nester1976}
F.W.~Hehl, P.~Von Der~Heyde, G.D.~Kerlick and J.M.~Nester, \emph{{General
  Relativity with Spin and Torsion: Foundations and Prospects}},
  \href{https://doi.org/10.1103/RevModPhys.48.393}{\emph{Rev. Mod. Phys.}
  {\bfseries 48} (1976) 393}.

\bibitem{Hayashi_Shirafuji1979}
K.~Hayashi and T.~Shirafuji, \emph{{New General Relativity}},
  \href{https://doi.org/10.1103/PhysRevD.19.3524}{\emph{Phys. Rev.} {\bfseries
  D19} (1979) 3524}.

\bibitem{Hayashi_Shirafuji1982}
K.~Hayashi and T.~Shirafuji, \emph{{Addendum to ``New General Relativity''}},
  {\emph{Phys. Rev.} {\bfseries D24} (1981) 3312}.

\bibitem{Bahamonde:2016kba}
S.~Bahamonde and C.G.~Böhmer, \emph{{Modified teleparallel theories of
  gravity: Gauss–Bonnet and trace extensions}},
  \href{https://doi.org/10.1140/epjc/s10052-016-4419-8}{\emph{The Eurpean
  Physical Journal} {\bfseries C76} (2016) 578}
  [\href{https://arxiv.org/abs/1606.05557}{{\ttfamily 1606.05557}}].

\bibitem{Obukhov:2006gea}
Y.N.~Obukhov, \emph{{Poincare gauge gravity: Selected topics}},
  \href{https://doi.org/10.1142/S021988780600103X}{\emph{Int. J. Geom. Meth.
  Mod. Phys.} {\bfseries 3} (2006) 95}
  [\href{https://arxiv.org/abs/gr-qc/0601090}{{\ttfamily gr-qc/0601090}}].

\bibitem{Fonseca-Neto:1992xln}
J.B.~Fonseca-Neto, M.J.~Rebou\c{c}as and A.F.F.~Teixeira, \emph{{The
  equivalence problem in torsion theories of gravitation}},
  \href{https://doi.org/10.1063/1.529577}{\emph{J. Math. Phys.} {\bfseries 33}
  (1992) 2574}.

\bibitem{Stephani:2003tm}
H.~Stephani, D.~Kramer, M.A.H.~MacCallum, C.~Hoenselaers and E.~Herlt,
  \emph{{Exact solutions of Einstein's field equations}}, Cambridge Monographs
  on Mathematical Physics, Cambridge Univ. Press, Cambridge (2003),
  \href{https://doi.org/10.1017/CBO9780511535185}{10.1017/CBO9780511535185}.

\bibitem{Coley:2022aty}
A.~Coley and R.~van~den Hoogen, \emph{{Teleparallel Geometry with a Single
  Affine Symmetry}}, \href{https://doi.org/10.1063/5.0099551}{\emph{J. Math.
  Phys.} {\bfseries 64} (2023) 022503}
  [\href{https://arxiv.org/abs/2205.07071}{{\ttfamily 2205.07071}}].

\bibitem{Maccallum:1972er}
M.A.H.~Maccallum and A.H.~Taub, \emph{{Variational principles and
  spatially-homogeneous universes, including rotation}},
  \href{https://doi.org/10.1007/BF01877589}{\emph{Commun. Math. Phys.}
  {\bfseries 25} (1972) 173}.

\end{thebibliography}\endgroup

\newpage
%
\appendix

\section{Appendix: Invariant Decomposition of the Torsion 2-form} \label{Torsion-Appendix}

In four dimensions the torsion 2-form can be decomposed into three irreducible parts invariant under the general linear group \cite{Hehl_McCrea_Mielke_Neeman1995}, called TENTOR ${}^{(1)}T^a$, TRATOR ${}^{(2)}T^a$, and AXITOR ${}^{(3)}T^a$, having $16$, $4$ and $4$ independent components respectively
\begin{equation}
T^a={}^{(1)}T^a+{}^{(2)}T^a+{}^{(3)}T^a.
\end{equation}
For a metric having Lorentzian signature the irreducible parts of the torsion are given by
\begin{eqnarray}
{}^{(2)}T^a&=&\frac{1}{3}h^a\wedge e_b\rfloor T^b,  \\
{}^{(3)}T^a&=&-\frac{1}{3}{}^*\!\left(h^a\wedge P\right), \quad\mathrm{with}\quad P={}^*\!\left(T^a\wedge h_a\right),\\
{}^{(1)}T^a&=&T^a-^{(2)}T^a-^{(3)}T^a.
\end{eqnarray}
The irreducible pieces satisfy the relations
\begin{eqnarray}
{}^{(1)}T^a\wedge h_a = 0 &\qquad&e_a\rfloor{}^{(1)}T^a=0\\
{}^{(2)}T^a\wedge h_a = 0 &\qquad&e_a\rfloor{}^{(3)}T^a=0
\end{eqnarray}

\section{Appendix: Contorsion} \label{contorsion appendix}

We introduce the antisymmetric contorsion 1-form $K^a_{\phantom{a}b}$ implicitly via the expression
\begin{equation}
T^a=K^a_{\phantom{a}b}\wedge h^b.
\end{equation}
If the nonmetricity $Q_{ab}=0$ then we find
\begin{equation}
\omega^a_{\phantom{a}b}=\omegaLC^a_{\phantom{a}b}+K^a_{\phantom{a}b}
\end{equation}
where $K_{ab}=-K_{ba}$.  We use an over circle to indicate the usual Riemannian (Levi-Civita) connection $\omegaLC^a_{\phantom{a}b}$ associated with the metric.
The contorsion one-form $K^a_{~b}$ can also be computed explicitly from the torsion two-form via the expression
\begin{equation}
K_{ab}=e_{[a} \rfloor T_{b]}-\frac{1}{2}(e_a\rfloor e_b \rfloor T_c)h^c.
\end{equation}

Interestingly, a relationship between the curvature of the spin connection $\omega^a_{\phantom{a}b}$, and the curvature of the Levi-Civita spin connection $\omegaLC^a_{\phantom{a}b}$ can be found
\begin{equation}
R^a_{\phantom{a}b}=\RLC^a_{\phantom{a}b} +\DLC K^a_{\phantom{a}b} + K^a_{\phantom{a}c}\wedge K^c_{\phantom{a}b},\label{Identity1}
\end{equation}
or as
\begin{equation}
R^a_{\phantom{a}b}=\RLC^a_{\phantom{a}b} +D K^a_{\phantom{a}b} - K^a_{\phantom{a}c}\wedge K^c_{\phantom{a}b}.\label{Identity2}
\end{equation}

The Ricci Scalar, $R$, can be computed via
\begin{equation}
R\,\eta = R^{ab}\eta_{ab}.
\end{equation}
It is now straight forward using equation \eqref{Identity2} to derive the following equivalent relationships
\begin{eqnarray}
R\,\eta &=& \RLC\,\eta+T\,\eta +d(K^{ab}\wedge \eta_{ab}) \\
R\,\eta &=& \RLC\,\eta+T\,\eta +d(h^{a}\wedge {}^*S_a) \\
R\,\eta &=& \RLC\,\eta+T\,\eta +d(2h^{a}\wedge{}^*T_a)
\end{eqnarray}
which in tele-parallel gravity when $R=0$, shows how the standard Hilbert Lagrangian of General Relativity $\RLC$ differs from the Lagrangian of $f(T)=T$ tele-parallel gravity by a total derivative.

\section{Appendix: Bianchi Geometries} \label{canonical-coordinates}

For a $G_3$ group of motions acting simply transitively on three dimensional spatial hypersurfaces, it is possible to express the spatial co-frame basis  $\mathcal{H}^I$, $I\in\{1,2,3\}$ with respect to some canonical coordinates $x^i=[x,y,z]$ \cite{Stephani:2003tm}. Given a basis of spatial co-frames $H^I$ for each Bianchi geometry one can construct the corresponding frame basis $\mathcal{E}_J$ determined via $\mathcal{E}_J\rfloor \mathcal{H}^I = \delta^I_J$.  The objects of anholonimity, $\mathcal{C}^I=d\mathcal{H}^I$,  are two-forms that can be invariantly decomposed as
\begin{eqnarray}
{}^{(2)}\mathcal{C}^I&=&\frac{1}{2}\, \mathcal{H}^I\wedge\left(\mathcal{E}_J \rfloor \mathcal{C}^J\right),\\
{}^{(1)}\mathcal{C}^I&=&\mathcal{C}^I-{}^{(2)}\mathcal{C}^I.
\end{eqnarray}
The three dimensional Bianchi geometries can broadly be divided into two classes.  The Class A Bianchi geometries are those in which ${}^{(2)}\mathcal{C}^I$ is identically zero, while the others belong to the Class B Bianchi geometries.

\subsection*{The Class A Bianchi Geometries}
For reference, the Class A geometries and their corresponding objects of anholonimity are:
\begin{description}
\item[Bianchi I]
\begin{eqnarray*}
\mathcal{H}^I &=& [dx, dy, dz] \\
\mathcal{C}^I &=& [0,0,0]\\
{}^{(2)}\mathcal{C}^I &=& [0,0,0]\\
{}^{(1)}\mathcal{C}^I &=&[0,0,0]
\end{eqnarray*}
\hrulefill
\item[Bianchi II]
\begin{eqnarray*}
\mathcal{H}^I &=& [dx-zdy, dy, dz] \\
\mathcal{C}^I &=& [\mathcal{H}^2\wedge \mathcal{H}^3,0,0]\\
{}^{(2)}\mathcal{C}^I &=& [0,0,0]\\
{}^{(1)}\mathcal{C}^I &=&[\mathcal{H}^2\wedge \mathcal{H}^3,0,0]
\end{eqnarray*}
\hrulefill
\item[Bianchi VI$_0$]
\begin{eqnarray*}
\mathcal{H}^I &=& [dx, (\cosh(x)dy-\sinh(x)dz), (-\sinh(x)dy+\cosh(x)dz)] \\
\mathcal{C}^I &=& [0, \mathcal{H}^3\wedge \mathcal{H}^1, -\mathcal{H}^1\wedge \mathcal{H}^2]\\
{}^{(2)}\mathcal{C}^I &=& [0,0,0]\\
{}^{(1)}\mathcal{C}^I &=&[0, \mathcal{H}^3\wedge \mathcal{H}^1, -\mathcal{H}^1\wedge \mathcal{H}^2]
\end{eqnarray*}
\hrulefill
\item[Bianchi VII$_0$]
\begin{eqnarray*}
\mathcal{H}^I &=& [dx, (\cos(x)dy-\sin(x)dz), (\sin(x)dy+\cos(x)dz)] \\
\mathcal{C}^I &=& [0, \mathcal{H}^3\wedge \mathcal{H}^1, \mathcal{H}^1\wedge \mathcal{H}^2]\\
{}^{(2)}\mathcal{C}^I &=& [0,0,0]\\
{}^{(1)}\mathcal{C}^I &=&[0,\mathcal{H}^3\wedge \mathcal{H}^1, \mathcal{H}^1\wedge \mathcal{H}^2]
\end{eqnarray*}
\hrulefill
\item[Bianchi VIII]
\begin{eqnarray*}
\mathcal{H}^I &=& [dx-\sinh(y) dz, (\cos(x)dy-\sin(x)\cosh(y)dz), (\sin(x)dy+\cos(x)\cosh(y)dz)] \\
\mathcal{C}^I &=& [-\mathcal{H}^2\wedge \mathcal{H}^3, \mathcal{H}^3\wedge \mathcal{H}^1, \mathcal{H}^1\wedge \mathcal{H}^2]\\
{}^{(2)}\mathcal{C}^I &=& [0,0,0]\\
{}^{(1)}\mathcal{C}^I &=&[-\mathcal{H}^2\wedge \mathcal{H}^3, \mathcal{H}^3\wedge \mathcal{H}^1, \mathcal{H}^1\wedge \mathcal{H}^2]
\end{eqnarray*}
\hrulefill
\item[Bianchi IX]
\begin{eqnarray*}
\mathcal{H}^I &=& [dx+\sin(y) dz, (\cos(x)dy-\sin(x)\cos(y)dz), (\sin(x)dy+\cos(x)\cos(y)dz)] \\
\mathcal{C}^I &=& [\mathcal{H}^2\wedge \mathcal{H}^3, \mathcal{H}^3\wedge \mathcal{H}^1, \mathcal{H}^1\wedge \mathcal{H}^2]\\
{}^{(2)}\mathcal{C}^I &=& [0,0,0]\\
{}^{(1)}\mathcal{C}^I &=&[\mathcal{H}^2\wedge \mathcal{H}^3, \mathcal{H}^3\wedge \mathcal{H}^1, \mathcal{H}^1\wedge \mathcal{H}^2]
\end{eqnarray*}
\hrulefill
\end{description}

\subsection*{The Class B Bianchi Geometries}
The Class B geometries and their corresponding objects of anholonimity are:
\begin{description}
\item[Bianchi IV]
\begin{eqnarray*}
\mathcal{H}^I &=& [dx, e^xdy, e^x(dz+xdy)] \\
\mathcal{C}^I &=& [0,\mathcal{H}^1\wedge \mathcal{H}^2,\mathcal{H}^1\wedge(\mathcal{H}^2+\mathcal{H}^3)]\\
{}^{(2)}\mathcal{C}^I &=& [0,\mathcal{H}^1\wedge \mathcal{H}^2,\mathcal{H}^1\wedge \mathcal{H}^3]\\
{}^{(1)}\mathcal{C}^I &=&[0,0,\mathcal{H}^1\wedge \mathcal{H}^2]
\end{eqnarray*}
\hrulefill
\item[Bianchi V]
\begin{eqnarray*}
\mathcal{H}^I &=& [dx, e^xdy, e^xdz] \\
\mathcal{C}^I &=& [0,\mathcal{H}^1\wedge \mathcal{H}^2,\mathcal{H}^1\wedge \mathcal{H}^3]\\
{}^{(2)}\mathcal{C}^I &=& [0,\mathcal{H}^1\wedge \mathcal{H}^2,\mathcal{H}^1\wedge \mathcal{H}^3]\\
{}^{(1)} \mathcal{C}^I &=&[0,0,0]
\end{eqnarray*}
\hrulefill
\item[Bianchi VI$_h$] where $h=-a^2$, $a\not = 0$ constant. (Bianchi $III$ when $a=1$.)
\begin{eqnarray*}
\mathcal{H}^I &=& [dx, e^{ax}(\cosh(x)dy-\sinh(x)dz), e^{ax}(-\sinh(x)dy+\cosh(x)dz)] \\
\mathcal{C}^I &=& [0,aH^1\wedge \mathcal{H}^2+\mathcal{H}^3\wedge \mathcal{H}^1,a\mathcal{H}^1\wedge \mathcal{H}^3-\mathcal{H}^1\wedge \mathcal{H}^2]\\
{}^{(2)}\mathcal{C}^I &=& [0,a\mathcal{H}^1\wedge \mathcal{H}^2,a\mathcal{H}^1\wedge \mathcal{H}^3]\\
{}^{(1)} \mathcal{C}^I &=& [0,\mathcal{H}^3\wedge \mathcal{H}^1,-\mathcal{H}^1\wedge \mathcal{H}^2]
\end{eqnarray*}
\hrulefill
\item[Bianchi VII$_h$] where $h=a^2$, $a\not = 0$ constant.
\begin{eqnarray*}
\mathcal{H}^I &=& [dx, e^{ax}(\cos(x)dy-\sin(x)dz), e^{ax}(\sin(x)dy+\cos(x)dz)] \\
\mathcal{C}^I &=& [0,a\mathcal{H}^1\wedge \mathcal{H}^2+\mathcal{H}^3\wedge \mathcal{H}^1,a\mathcal{H}^1\wedge \mathcal{H}^3-\mathcal{H}^1\wedge \mathcal{H}^2]\\
{}^{(2)}\mathcal{C}^I &=& [0,a\mathcal{H}^1\wedge \mathcal{H}^2,a\mathcal{H}^1\wedge \mathcal{H}^3]\\
{}^{(1)} \mathcal{C}^I &=& [0,\mathcal{H}^3\wedge \mathcal{H}^1,\mathcal{H}^1\wedge \mathcal{H}^2]
\end{eqnarray*}
\hrulefill
\end{description}

\end{document}